\documentclass[aps,prd,showpacs,notitlepage,nofootinbib,superscriptaddress,floatfix,showkeys,twocolumn]{revtex4-2}
\usepackage{blindtext}
\usepackage{hyperref}
\usepackage{amsmath,amssymb}
\usepackage{float}
\usepackage{microtype}
\usepackage{marginnote}
\usepackage{graphicx}
\usepackage{bm}
\usepackage{latexsym}
\usepackage{epsfig}
\usepackage{psfrag}
\usepackage{color}
\usepackage[dvipsnames]{xcolor}
\usepackage{subfigure}
\usepackage{graphicx} 
\usepackage{amssymb}
\usepackage{amsmath}
\usepackage{bm}
\usepackage{latexsym}
\usepackage{epsfig}
\usepackage{psfrag}
\usepackage[normalem]{ulem}
\usepackage{textcomp}
\usepackage{color}
\usepackage{pstricks}
\usepackage[utf8]{inputenc}
\usepackage{comment}
\usepackage{braket}

\def\nn{\nonumber} 
\def\pa{{\partial}}
\def\f{\frac}
\def\l{\left}
\def\r{\right}
\def\d{{\mathrm {d}}}

\def\pa{\partial}
\def\Mpl{M_{_{\mathrm{Pl}}}}

\def\cA{{\mathcal{A}}}

\def\cG{{\mathcal G}}
\def\cR{{\mathcal R}}
\def\cB{{\mathcal B}}
\def\ps{\mathcal{P}_{_{\mathrm{S}}}}
\def\pb{\mathcal{P}_{_{\mathrm{B}}}}
\def\ns{n_{_{\mathrm{S}}}}
\def\pbm{\mathcal{P}_{_{\mathrm{B}}}^{_{\mathrm{M}}}}

\def\ei{\eta_{\mathrm{i}}}
\def\ee{\eta_{\mathrm{e}}}
\def\ai{a_{\mathrm{i}}}
\def\e1i{\epsilon_{1\mathrm{i}}}

\def\HI{H_{_{\mathrm{I}}}}
\def\bnl{b_{_{\mathrm{NL}}}}

\def\vx{{\bm{x}}}
\def\vk{{\bm{k}}}

\def\ps{\mathcal{P}_{_{\mathrm{S}}}}
\def\pt{\mathcal{P}_{_{\mathrm{T}}}}
\def\rb{\rho_{_{\mathrm{B}}}}
\def\re{\rho_{_{\mathrm{E}}}}
\def\rem{\rho_{_{\mathrm{EM}}}}
\def\pb{\mathcal{P}_{_{\mathrm{B}}}}
\def\pe{\mathcal{P}_{_{\mathrm{E}}}}
\def\nb{n_{_{\mathrm{B}}}}
\def\ne{n_{_{\mathrm{E}}}}

\def\ee{\eta_{\mathrm{e}}}
\def\eai{\epsilon_{1\mathrm{i}}}
\def\ai{a_{\mathrm{i}}}

\def\mpcinv{{\rm Mpc}^{-1}}

\def\fnl{f_{_{\mathrm{NL}}}}

\allowdisplaybreaks[1]

\begin{document}
%%%%%%%%%%%%%%%%%%%%%%%%%%%%%%%%%%%%%%%%%%%%%%%%%%%%%%%%%%%%%%%%%%%%%%%%%%%%%%%
\title{Cross-correlation between the curvature perturbations\\ 
and magnetic fields in {\it pure}\/ ultra slow roll inflation}
\author{Sagarika Tripathy}
\email{E-mail: sagarika.tripathy@iiap.res.in} 
\affiliation{Indian Institute of Astrophysics, II Block, Koramangala, 
Bengaluru 560034, India}
\author{Debika Chowdhury}
\email{E-mail: debika.chowdhury@iiap.res.in}
\affiliation{Indian Institute of Astrophysics, II Block, Koramangala, 
Bengaluru 560034, India}
\author{H.~V.~Ragavendra}
\email{E-mail: ragavendra.hv@pd.infn.it}
\affiliation{Raman Research Institute, C.~V.~Raman Avenue, Sadashivanagar, 
Bengaluru 560080, India}
\affiliation{Dipartimento di Fisica e Astronomia “Galileo Galilei”, Universit\`{a} degli Studi di Padova, Via Marzolo 8, I-35131 Padova, Italy}
\affiliation{Istituto Nazionale di Fisica Nucleare, Sezione di Padova, Via Marzolo 8, I-35131 Padova, Italy}
\author{L.~Sriramkumar}
\email{E-mail: sriram@physics.iitm.ac.in}
\affiliation{Centre for Strings, Gravitation and Cosmology,
Department of Physics, Indian Institute of Technology Madras, 
Chennai~600036, India}
%%%%%%%%%%%%%%%%%%%%%%%%%%%%%%%%%%%%%%%%%%%%%%%%%%%%%%%%%%%%%%%%%%%%%%%%%%%%%%%
\begin{abstract}
Motivated by the aim of producing significant number of primordial black
holes, over the past few years, there has been a considerable interest in 
examining models of inflation involving a single, canonical field, that 
permit a brief period of ultra slow roll. 
Earlier, we had examined inflationary magnetogenesis---achieved by
breaking the conformal invariance of the electromagnetic action through
a coupling to the inflaton---in situations involving departures from
slow roll.
We had found that a transition from slow roll to ultra slow roll inflation
can lead to a strong blue tilt in the spectrum of the magnetic field over 
small scales and also considerably suppress its strength over large scales. 
In this work, we consider the scenario of {\it pure}\/ ultra slow roll
inflation and show that scale invariant magnetic fields can be obtained
in such situations with the aid of a non-conformal coupling function 
that depends on the kinetic energy of the inflaton. 
Apart from the power spectrum, an important probe of the primordial 
magnetic fields are the three-point functions, specifically, the
cross-correlation between the curvature perturbations and the magnetic
fields. 
We calculate the three-point cross-correlation between the curvature
perturbations and the magnetic fields in {\it pure}\/ ultra slow roll 
inflation for the new choice of the non-conformal coupling function. 
In particular, we examine the validity of the consistency condition that 
is expected to govern the three-point function in the squeezed limit and
comment on the wider implications of the results we obtain.
\end{abstract}
\maketitle

%%%%%%%%%%%%%%%%%%%%%%%%%%%%%%%%%%%%%%%%%%%%%%%%%%%%%%%%%%%%%%%%%%%%%%%%%%%%%%%

\section{Introduction}

Large-scale, coherent magnetic fields are observed in galaxies, clusters of
galaxies, and even in the intergalactic medium (for reviews on magnetic fields, 
see Refs.~\cite{Grasso:2000wj,Giovannini:2003yn,Brandenburg:2004jv,Kulsrud:2007an,
Subramanian:2009fu,Kandus:2010nw,Widrow:2011hs,Durrer:2013pga,Subramanian:2015lua,
Vachaspati:2020blt,Vachaspati:2020blt}). 
The Fermi/LAT and HESS observations of TeV blazars suggest that the
strength of the magnetic fields in the intergalactic voids is of the
order of $10^{-16}\,\mathrm{G}$~\cite{Neronov:1900zz,Tavecchio:2010mk,
Dolag:2010ni,Dermer:2010mm,Vovk:2011aa,Taylor:2011bn, Takahashi:2011ac}.
It may not be possible to explain the presence of magnetic fields in the 
intergalactic voids solely on the basis of astrophysical phenomena such 
as batteries (see Refs.~\cite{Brandenburg:2004jv,Kulsrud:2007an};
however, in this context, see Refs.~\cite{Papanikolaou:2023nkx,
Papanikolaou:2023cku}).
Hence, it seems necessary to invoke a cosmological mechanism for the 
primordial origin of these fields.
While the Fermi/LAT and HESS observations provide a lower bound on the 
current strengths of the cosmological magnetic fields, the anisotropies 
in the cosmic microwave background (CMB) provide an upper bound.
The primordial magnetic fields can induce scalar, vector, and tensor 
perturbations, which, in turn, can leave distinct signatures on the 
temperature and polarization angular power spectra of the 
CMB~\cite{Durrer:1999bk,Giovannini:2007qn,Finelli:2008xh,Paoletti:2008ck,
Shaw:2009nf,Bonvin:2013tba,Ballardini:2014jta}.
The CMB observations lead to an upper bound on the primordial magnetic 
fields to be of the order of $10^{-9}\,\mathrm{G}$~\cite{Shaw:2010ea,
Paoletti:2010rx,Planck:2015zrl,Zucca:2016iur,Paoletti:2018uic,Paoletti:2019pdi,
Paoletti:2022gsn}.

As in the case of the scalar and tensor perturbations, it is natural to 
turn to the inflationary epoch in the early universe for the generation 
of magnetic fields.
However, since the standard electromagnetic theory is conformally invariant,
the magnetic fields generated in such a case will have a strongly blue
spectrum, whose strength will be significantly suppressed during inflation. 
Therefore, to generate cosmological magnetic fields of observable strengths
today, the conformal invariance of the electromagnetic action must be broken
during inflation (see Refs.~\cite{Turner:1987bw,Ratra:1991bn};
for reviews in this context, see Refs.~\cite{Subramanian:2009fu,
Kandus:2010nw,Durrer:2013pga,Subramanian:2015lua,Vachaspati:2020blt}).
Often, this is achieved by the coupling the electromagnetic field to the 
inflaton, i.e. the scalar field(s) that drive inflation~\cite{Bamba:2003av,
Martin:2007ue,Watanabe:2009ct,Kanno:2009ei,Markkanen:2017kmy}.

Since the detection of gravitational waves from merging binary black 
holes~\cite{LIGOScientific:2020kqk}, there has been a tremendous 
interest in investigating if these black holes could have formed in 
the primeval universe~\cite{DeLuca:2020qqa,Jedamzik:2020ypm,
Jedamzik:2020omx,Franciolini:2021tla}.
If a significant number of black holes have to be produced in the early
universe, then the primordial scalar power spectrum should have considerably
higher power on small scales than the nearly scale invariant amplitude
suggested by the anisotropies in the CMB over the large scales.
In single field models of inflation involving the canonical scalar field,
scalar spectra with enhanced power on small scales can be generated if the 
models admit a brief period of ultra slow roll during which the first slow 
roll parameter decreases very rapidly.
While the first slow roll parameter remains small during the period of ultra
slow roll, the second and higher order slow roll parameters turn large 
suggesting strong departures from slow roll inflation.
It is known that such deviations from slow roll lead to enhanced levels of
non-Gaussianities, in particular, the three-point functions (in this context,
see, for example, Refs.~\cite{Hazra:2012yn,Sreenath:2014nca,Atal:2018neu,
Atal:2019erb,Ragavendra:2020sop,Ragavendra:2021qdu,Cai:2022erk,
Ragavendra:2023ret,Namjoo:2024ufv}).

We had mentioned above that, to generate magnetic fields of observable
strengths, during inflation, the conformal invariance of the electromagnetic
action is usually broken by introducing a coupling to the inflaton.
Recently, we had shown that, in single field models of inflation admitting
a phase of ultra slow roll, there arises a challenge in generating magnetic
fields of observable strengths over large scales~\cite{Tripathy:2021sfb,
Tripathy:2022iev}.
The phase of ultra slow roll is typically achieved with the help of a point
of inflection in the inflationary potential.
The scalar field slows down tremendously as it approaches the point of 
inflection, leading to the period of ultra slow roll.
As a result, after the onset of ultra slow roll, the non-conformal coupling
function ceases to evolve, essentially restoring the conformal invariance of
the electromagnetic action.
Such a behavior leads to a strongly scale dependent spectrum of the magnetic
field on small scales.
Also, the amplitude of the spectrum of the magnetic field is considerably
suppressed over large scales.

In this work, we shall examine the effects of ultra slow roll inflation on
the three-point cross-correlation between the curvature perturbations and
the magnetic fields.
Specifically, we shall focus on the scenario wherein ultra slow roll lasts 
for the {\it entire}\/ duration of inflation, which we shall refer to as 
{\it pure}\/ ultra slow roll inflation.
Although such a scenario may not be considered as interesting from an
observational point of view, the motivations for our investigations are
twofold.
Firstly, our aim will be to examine if we can construct suitable non-conformal
coupling functions that can lead to a scale invariant spectrum for the magnetic
field in ultra slow roll inflation.
Secondly, in slow roll inflation, it is well known that there arises a consistency
condition according to which the three-point functions involving the scalar and
tensor perturbations can be completely expressed in terms of the two-point functions
in the so-called squeezed limit wherein one of the three wave numbers is much
smaller than the other two.
This property can be attributed to the fact that the amplitude of the scalar
and tensor perturbations freeze on super-Hubble scales.
However, it is known that the strength of the curvature perturbations can grow
indefinitely on super-Hubble scales when there arises an extended period of
ultra slow roll.
Due to the continued growth in the amplitude of the curvature perturbations,
it has been shown that, in {\it pure}\/ ultra slow roll inflation, the
standard consistency condition governing the scalar bispectrum is
violated~\cite{Namjoo:2012aa,Martin:2012pe,Suyama:2020akr,Bravo:2020hde,
Suyama:2021adn}.
Such a result raises the interesting question of whether, in similar situations, 
the three-point cross-correlation between the curvature perturbations and the
magnetic field satisfies the expected consistency condition.

This manuscript is organized as follows.
In the following section, we shall discuss the behavior of the background 
and the Fourier mode functions describing the curvature perturbations in 
specific {\it pure}\/ ultra slow roll scenarios.
We shall also outline the generation of electromagnetic fields by a 
non-conformal coupling function that depends on the inflaton and
discuss the behavior of the Fourier mode functions describing the 
electromagnetic vector potential.
Further, we shall arrive at the power spectra of the curvature perturbations 
and the electromagnetic fields in these cases.
In Sec.~\ref{sec:JX}, we shall consider a non-conformal coupling function 
that depends on the kinetic energy of the inflaton.
We shall show that, for suitable choices of the parameters involved, such 
a coupling function leads to the desired scale invariant spectrum for the 
magnetic field in the ultra slow roll scenarios. 
In Sec.~\ref{sec:RBB}, we shall derive the third order action that describes 
the interaction between the curvature perturbation and the electromagnetic 
field for the new choice of the non-conformal coupling function. 
We shall also arrive at formal expressions for the contributions to the 
three-point cross-correlation (in Fourier space) between the curvature 
perturbations and the magnetic fields corresponding to the different terms 
in the action at the cubic order.
We shall also introduce the non-Gaussianity parameter associated with the 
three-point cross-correlation of interest.
In Sec.~\ref{sec:calB}, we shall calculate the three-point cross-correlation 
for two specific cases of pure ultra slow roll inflation and suitable 
values of the relevant parameters. 
We shall also illustrate the complete structure of the non-Gaussianity parameter
and discuss the validity of the consistency condition governing the parameter in 
the scenarios we consider.
Lastly, in Sec.~\ref{sec:conc}, we shall conclude with a summary and outlook.
We relegate complementary details of the computation and arguments on certain
related issues to the appendices.

At this stage of our discussion, we should make a few clarifying remarks
concerning the conventions and notations that we shall work with.
We shall work with natural units such that $\hbar=c=1$ and set the reduced 
Planck mass to be $\Mpl=\l(8\,\pi\, G\r)^{-1/2}$.
We shall adopt the signature of the metric to be~$(-,+,+,+)$.
Note that Latin indices shall represent the spatial coordinates, except 
for~$k$ which shall be reserved for denoting the wave number. 
We shall assume the background to be the spatially flat
Friedmann-Lema\^itre-Robertson-Walker~(FLRW) universe described by the 
following line-element:
\begin{equation}
\d s^2=-\d t^2+a^2(t)\,\d {\bm x}^2,\label{eq:flrw-le}
\end{equation}
where~$t$ is the cosmic time and~$a(t)$ denotes the scale factor.
Also, an overdot and an overprime shall denote differentiation with respect to 
the cosmic time~$t$ and the conformal time~$\eta=\int \d t/a(t)$, respectively.

%%%%%%%%%%%%%%%%%%%%%%%%%%%%%%%%%%%%%%%%%%%%%%%%%%%%%%%%%%%%%%%%%%%%%%%%%%%%%%%

\section{Background, mode functions and power spectra}

In this section, we shall discuss the behavior of the background as well as 
the Fourier mode functions of the curvature perturbation and the electromagnetic
field in ultra slow roll inflation.
We shall also discuss the power spectra of the curvature perturbations and
the electromagnetic fields that arise in a couple of specific situations.

%%%%%%%%%%%%%%%%%%%%%%%%%%%%%%%%%%%%%%%%%%%%%%%%%%%%%%%%%%%%%%%%%%%%%%%%%%%%%%%

\subsection{Evolution of the background}

Recall that the equation of motion governing a homogeneous, canonical 
scalar field, say, $\phi$, that is described by the potential $V(\phi)$ 
is given by
\begin{equation}
\ddot{\phi}+3\,H\,\dot{\phi}+V_\phi=0,
\end{equation}
where $H=\dot{a}/a$ is the Hubble parameter and $V_\phi=\d V/\d \phi$.
As we mentioned above, a brief epoch of ultra slow roll inflation is 
usually realized in potentials that contain a point of inflection, i.e.
a point wherein the first and the second derivatives of the potential,
viz. $V_{\phi}$ and $V_{\phi\phi}=\d^2V/\d \phi^2$, vanish (see, for 
instance, Refs.~\cite{Germani:2017bcs,Bhaumik:2019tvl,Ragavendra:2020sop}).
Note that, when $V_\phi=0$, the equation of motion for the scalar field
reduces to
\begin{equation}
\ddot{\phi}=-3\,H\,\dot{\phi},\label{eq:vVphi}
\end{equation}
which can be immediately integrated to obtain that $\dot{\phi} \propto a^{-3}$. 
Also, the fact that $V_\phi=0$ suggests that the Hubble parameter $H$ is 
nearly a constant.
In such a situation, the first slow roll parameter $\epsilon_1=\dot{\phi}^2/(2\,
H^2\,\Mpl^2)$ behaves as
\begin{equation}
\epsilon_1\propto\f{1}{a^6}.
\end{equation}

In this work, we shall be interested in {\it pure}\/ ultra slow roll inflation, 
i.e. scenarios that lead to ultra slow roll for the {\it entire}\/ duration of
inflation.
To allow for different types of evolution of the first slow parameter, we 
shall modify Eq.~\eqref{eq:vVphi} and consider an equation of the form (for 
a detailed discussion in this regard, see Refs.~\cite{Martin:2012pe,
Motohashi:2017aob,Motohashi:2014ppa})
\begin{equation}
\ddot{\phi}=-\f{p}{2}\,H\,\dot{\phi},
\end{equation}
with $p=6$ evidently corresponding to the original case.
Clearly, this leads to $\dot{\phi} \propto a^{-p/2}$ and, for $p>0$, the 
velocity as well as the acceleration of the field continue to decrease.
If we assume that the initial value of the first slow roll parameter is small
(say, $\epsilon_{1\mathrm{i}} \lesssim 10^{-2}$), since it decreases rapidly
thereafter, the Hubble parameter remains nearly a constant.
Hence, the scale factor during ultra slow roll inflation can be expressed in 
the de Sitter form as $a(\eta)=-1/(\HI\,\eta)$, where $\HI$ denotes the 
constant value of the Hubble parameter.
Therefore, in such situations, we can express the first slow roll parameter as
\begin{equation}
\epsilon_1(\eta) = \e1i\,\l[\f{\ai}{a(\eta)}\r]^p
=\eai\,\l(\f{\eta}{\ei}\r)^{p},\label{eq:e1-p}
\end{equation}
where~$\eai$ is the initial value of the first slow roll parameter when the 
value of the scale factor is~$\ai=-1/(\HI\,\ei)$ at the conformal time~$\ei$.
In our discussion below, we shall consider two cases wherein $p=6$ and $p=4$.
As we shall see, the first of these leads to a scale invariant spectrum for the
curvature perturbations, while the second leads to simple mode functions, which
allows us to easily calculate the two and three-point functions of 
interest\footnote{We should clarify a point on nomenclature here.
We find that scenarios with constant~$p$ are also referred to as constant
roll inflation~\cite{Motohashi:2014ppa,Motohashi:2017aob}.
We shall be focusing on scenarios wherein $p>3$. 
In such cases, the first slow roll parameter decreases so rapidly that
the amplitude of the curvature perturbations grows indefinitely at late
times.
Hence, we believe it is more apt to refer to these scenarios as ultra slow
roll inflation.}.

%%%%%%%%%%%%%%%%%%%%%%%%%%%%%%%%%%%%%%%%%%%%%%%%%%%%%%%%%%%%%%%%%%%%%%%%%%%%%%%

\subsection{Mode functions describing the curvature perturbation and 
the scalar power spectrum}

Let us now discuss the mode functions describing the curvature perturbation
and the resulting scalar power spectrum in the ultra slow roll scenarios of
interest.
Let~$\cR$ denote the curvature perturbation and let~$f_k$ denote the corresponding
Fourier mode functions.
Recall that the Fourier mode functions~$f_k$ that characterize the curvature 
perturbation satisfy the differential equation (see, for 
example, the reviews~\cite{Mukhanov:1990me,Martin:2003bt,Martin:2004um,
Bassett:2005xm,Sriramkumar:2009kg,Baumann:2008bn,Baumann:2009ds,
Sriramkumar:2012mik,Linde:2014nna,Martin:2015dha})
\begin{eqnarray}
f_{k}''+2\,\f{z'}{z}\,f_{k}+k^2\,f_{k}=0, \label{eq:F-Scp}
\end{eqnarray}
where $z= \sqrt{2\,\epsilon_1}\,\Mpl\,a$. 
On quantization, the curvature perturbation $\cR$ can be expressed 
in terms of the mode functions~$f_k$ as follows:
\begin{align}
\hat{\cR}(\eta,{\bm x})
&= \int\f{\d^3{\bm k}}{(2\,\pi)^{3/2}}\, 
\hat{\cR}_{\bm k}(\eta)\,\mathrm{e}^{i\,\bm{k}\cdot{\bm x}} \nn\\
&= \int\f{\d^3{\bm k}}{(2\,\pi)^{3/2}}\, 
\l[\hat{a}_{\bm k}\,f_{k}(\eta)\,\mathrm{e}^{i\,{\bm k}\cdot{\bm x}} 
+ \hat{a}_{\bm{k}}^{\dag}\, f_{k}^*(\eta)\,
\mathrm{e}^{-i\,{\bm k}\cdot\bm{x}}\r],
\end{align}
where the annihilation and creation operators $\hat{a}_{\bm k}$ 
and~$\hat{a}_{\bm k}^{\dag}$ satisfy the standard commutation 
relations, viz.
\begin{equation}
[\hat{a}_{\bm k},\hat{a}_{{\bm k}'}]
=[\hat{a}_{\bm k}^\dag,\hat{a}_{{\bm k}'}^\dag]=0,\;
[\hat{a}_{\bm k},\hat{a}_{{\bm k}'}^\dag]
=\delta^{(3)}\l({\bm k}-{\bm k}'\r).
\end{equation}
To obtain the solutions for the mode functions~$f_k$, one often 
works with the associated Mukhanov-Sasaki variable~$v_k$---defined 
as $v_k=z\,f_k$---which satisfies the differential equation
\begin{equation}
v_k'' + \l(k^2 - \f{z''}{z} \r)\,v_k= 0. \label{eq:de-vk} 
\end{equation}
The scalar power spectrum is defined as
\begin{equation}
\ps(k) = \f{k^3}{2\, \pi^2}\,\vert f_k\vert^2
=\f{k^3}{2\, \pi^2}\,\f{\vert v_k\vert^2}{z^2},\label{eq:ps-d}
 \end{equation}
where the quantities $f_k$, $v_k$ and $z$ are to be evaluated at 
late times close to the end of inflation.

Let us first consider the case of ultra slow inflation wherein~$p=6$.
In such a case, we have $\epsilon_1=\epsilon_{1\mathrm{i}}\,(\ai/a)^6$
and $z=\sqrt{2\,\eai}\,\Mpl\, \ai^3\,(\HI\,  \eta)^2$, so that 
$z''/z=2/\eta^2$.
As this is similar to the case of pure de Sitter, the solution to the 
mode function~$f_k$, which satisfies the standard Bunch-Davies condition 
in the domain wherein $k \gg \sqrt{z''/z}$ [which corresponds to the 
sub-Hubble limit $k/(a\,H)\simeq (-k\,\eta) \gg 1$], can be immediately 
obtained to be
\begin{align}
f_k(\eta) &= \f{1}{\sqrt{2\,\eai}\,\Mpl\,\ai^3\,\HI^2\,\eta^2}\,
\f{1}{\sqrt{2\,k}}\,\l(1-\f{i}{k\,\eta}\r)\, \mathrm{e}^{-i\,k\,\eta}\nn\\
&=\f{-\HI}{\sqrt{4\,k\,\eai}\,\Mpl}\,\l(\f{\ei^3}{\eta^2}\r)\,
\l(1-\f{i}{k\,\eta}\r)\, \mathrm{e}^{-i\,k\,\eta}.\label{eq:fk-p6}
\end{align}
The resulting scalar power spectrum, evaluated at a late time, say,~$\ee$,
corresponding to the end of inflation, turns out to be
\begin{equation}
\ps(k) = \f{\HI^2}{8\, \pi^2\,  \Mpl^2\, \eai}\,\l[\f{a(\ee)}{\ai}\r]^6
=\f{\HI^2}{8\, \pi^2\,  \Mpl^2\, \eai}\,\l(\f{\ei}{\ee}\r)^6.
\label{eq:ps-p6}
\end{equation}
There are two points concerning this scalar power spectrum that need to 
be emphasized.
The first point is the fact the spectrum is independent of~$k$, i.e. it is 
scale invariant.
The second point is that, in contrast to slow roll inflation, the spectrum
does not approach a constant value at late times [i.e.~when $k \ll \sqrt{z''/z}$,
which is equivalent to the super-Hubble limit $k/(a\,H)\simeq (-k\,\eta) \ll 1$].
In fact, the power spectrum grows as~$a^6$ at late times. 
This can be attributed to the fact that the mode function $f_k$ grows as 
$a^3$ in the super-Hubble limit.
Such a growth is a well known feature in the ultra slow roll inflationary 
scenarios~\cite{Namjoo:2012aa,Martin:2012pe}.

Let us now turn to the case of $p=4$.
When $p=4$, we have $\epsilon_1=\eai\,(\ai/a)^4$ and $z=\sqrt{2\,\eai}\,
\Mpl\,(-\HI\,\eta)$ so that $z''/z=0$. 
It should be evident that the solution to the mode function~$f_k$ which
satisfies the Bunch-Davies initial condition is given by 
\begin{align}
f_{k}(\eta) 
&= \f{1}{{\sqrt{2\, \eai}\,\ai^2\,\Mpl\,(-\HI\,\eta)}}
\f{1}{\sqrt{2\,k}}\,\mathrm{e}^{-i\,k\,\eta} \nn \\
&= \f{-\HI}{\sqrt{4\,k\,\eai}\,\Mpl}\,\l(\f{\ei^2}{\eta}\r)\,
\mathrm{e}^{-i\,k\,\eta}.\label{eq:fk-p4}
\end{align}
The scalar power spectrum evaluated at the end of inflation (corresponding 
to the conformal time~$\ee$) can be expressed as
\begin{align}
\ps(k) &= \f{\HI^2}{8\,\pi^2\,\Mpl^2\,\eai}\,\l(\f{k}{\ai\,\HI}\r)^2\,
\l[\f{a(\ee)}{\ai}\r]^2\nn\\
&=\f{\HI^2}{8\,\pi^2\,\Mpl^2\,\eai}\,\l(\f{\ei}{\ee}\r)^2
\l(k\,\ei\r)^2.\label{eq:ps-p4}
\end{align}
As in the case of $p=6$, the mode function $f_k$ and the scalar power 
spectrum $\ps(k)$ grow indefinitely at late times.
Moreover, note that, the above spectrum has a strong blue tilt, as it 
behaves as~$k^2$.

%%%%%%%%%%%%%%%%%%%%%%%%%%%%%%%%%%%%%%%%%%%%%%%%%%%%%%%%%%%%%%%%%%%%%%%%%%%%%%%

\subsection{Mode functions describing the electromagnetic field 
and power spectra}\label{subsec:PBPE}

In a FLRW universe, the conformal invariance of the standard electromagnetic 
action dilutes the amplitude of the cosmological magnetic fields as $B\propto 
1/a^2$.
The dilution is, in particular, considerable during the inflationary epoch.
Therefore, to generate magnetic fields during inflation that are consistent 
with the current observations (say, $10^{-16} < B_0 < 10^{-9}\,\mathrm{G}$), 
the conformal invariance of the electromagnetic action must be broken. 
This is usually achieved by introducing a function in the action which couples 
the electromagnetic field to the inflaton, say, $\phi$.
The electromagnetic action that is often considered in such a case has the 
form (see, for example, Refs.~\cite{Martin:2007ue,Subramanian:2009fu})
\begin{equation}
S[A^\mu] = -\f{1}{16\,\pi} \int \d^4x \sqrt{-g}\, J^2(\phi)\, 
F_{\mu\nu}\,F^{\mu\nu},\label{eq:ema-1}
\end{equation}
where $J(\phi)$ denotes the non-conformal coupling function and the field 
tensor~$F_{\mu\nu}$ is expressed in terms of the vector potential~$A_\mu$ 
as~$F_{\mu\nu}=( \pa_{\mu}\,A_{\nu}-\pa_{\nu}\,A_{\mu})$.

We can arrive at the equation of motion governing the vector potential $A_{\mu}$
by varying the above action.
In what follows, we shall choose to work in the Coulomb gauge wherein $A_0 
= 0$ and $(g^{ij}\,\partial_i A_j)=0$. 
Let $\bar{A}_k$ denote the Fourier mode functions associated with the vector potential~$A_i$.
We find that the mode functions $\bar{A}_k$ satisfy the following differential
equation (see, for example, Refs.~\cite{Martin:2007ue,Subramanian:2009fu}):
\begin{equation}
\bar{A}_k''+2\, \f{J'}{J}\,\bar{A}_k' +k^2 \bar{A}_k = 0. 
\label{eq:de-Abk}
\end{equation}
If we write $\bar{A}_k=\cA_k/J$, then this differential equation simplifies 
to the form
\begin{equation}
\cA_k''+ \l(k^2- \f{J''}{J}\r)\, \cA_k = 0.\label{eq:de-cAk}
\end{equation}
For each comoving wave vector ${\bm k}$, we can define the polarization vector
${\bm \varepsilon}_{\lambda}^{\bm k}$, where $\lambda=\{1,2\}$ represents the 
two states of polarization of the electromagnetic field.
The polarization vector satisfies the 
condition ${\bm k}\cdot{\bm \varepsilon}_{\lambda}^{\bm k}=0$.
Moreover, the components $\varepsilon_{\lambda i}^{\bm k}$ of the polarization
vector satisfy the relation
\begin{equation}
\sum_{\lambda=1}^2 \varepsilon^{\bm k}_{\lambda i}\,
\varepsilon^{\bm k}_{\lambda j} 
=\delta_{ij}-\f{k_i\,k_j}{k^2},
\end{equation}
where $k_i$ denotes the components of the wave vector~${\bm k}$.
On quantization, the vector potential $A_i$ can be decomposed in terms of 
the mode functions~$\bar{A}_k$ as follows:
\begin{align}
\hat{A}_i(\eta,{\bm x}) 
&= \sqrt{4\,\pi}\int  \f{\d^3{\bm k}}{(2\,\pi)^{3/2}} 
\sum_{\lambda=1}^{2} \varepsilon_{\lambda i}^{\bm k}\,
\biggl[\hat{b}_{\bm k}^{\lambda}\, 
\bar{A}_k(\eta)\,\mathrm{e}^{i\,{\bm k} \cdot{\bm x}}\nn\\
& + \hat{b}_{\bm k}^{\lambda \dag}\, \bar{A}_k^{\ast}(\eta) 
\mathrm{e}^{-i\,{\bm k} \cdot{\bm x}}\biggr].\label{eq:fAk}
\end{align}
The creation and annihilation operators $\hat{b}_{\bm k}^{\lambda}$ and 
$\hat{b}_{\bm k}^{\lambda\dag}$ satisfy the following commutation relations:
\begin{equation}
[\hat{b}_{\bm k}^{\lambda},\hat{b}_{\bm k'}^{\lambda'}]
=[\hat{b}_{\bm k}^{\lambda\dag},\hat{b}_{\bm k'}^{\lambda'\dag}] =0,\;
[\hat{b}_{\bm k}^{\lambda},\hat{b}_{\bm k'}^{\lambda'\dag}]
= \delta^{(3)}({\bm k}- {\bm k'})\,\delta^{\lambda\lambda'}.
\end{equation}

The energy densities associated with the electric and magnetic fields, 
say, $\rb$ and~$\re$, are given by 
\begin{subequations}\label{eq:rbe}
\begin{align}
\rb &= \frac{J^2}{16\, \pi\, a^4}\,\delta^{im}\,\delta^{jn}\,F_{ij}\,F_{mn},\\
\re &= \f{J^2}{8\, \pi\, a^4}\,\delta^{ij}\,A_i'\,A_j'.
\end{align}
\end{subequations}
Let $\hat{\rho}_{_{\mathrm{B}}}$ and $\hat{\rho}_{_{\mathrm{E}}}$ denote
the operators associated with the corresponding energy densities.
The power spectra of the magnetic and electric fields, say, $\pb(k)$ 
and~$\pe(k)$, are defined as~\cite{Martin:2007ue,Subramanian:2009fu}
\begin{equation}
\pb(k) = \f{\d \langle 0\vert \hat{\rho}_{_{\mathrm{B}}} \vert 0\rangle}{\d\, 
\mathrm{ln}\, k},\;
\pe(k) = \f{\d \langle 0\vert \hat{\rho}_{\mathrm{E}}
\vert 0\rangle}{\d\, \mathrm{ln}\, k},\label{eq:pbe-d}
\end{equation}
where $\vert 0\rangle$ denotes the vacuum state annihilated by the 
annihilation operator $\hat{b}_{\vk}^\lambda$, i.e. $\hat{b}_{\vk}^\lambda
\vert 0\rangle=0$ for all $\vk$ and $\lambda$.
On utilizing the decomposition of the vector potential in Eq.~\eqref{eq:fAk},
the power spectra $\pb(k)$ and~$\pe(k)$ can be expressed in terms of 
the mode functions~$\bar{A}_k$ and their scaled forms~$\cA_k$ as 
follows~\cite{Martin:2007ue,Subramanian:2009fu}:
\begin{subequations}\label{eq:pbe}
\begin{align}
\pb(k) &= \f{k^5}{2\,\pi^2}\,\f{J^2}{a^4}\, 
\vert \bar{A}_k\vert^2
=\f{k^5}{2\,\pi^2\, a^4}\, \vert \cA_k\vert^2,\label{eq:pb}\\ 
\pe(k) &= \f{k^3}{2\,\pi^2}\,\,\f{J^2}{a^4}\, \vert \bar{A}_k'\vert^2
= \f{k^3}{2\,\pi^2\, a^4}\, \biggl\vert \cA_k'
-\f{J'}{J}\,\cA_k\biggr\vert^2.\quad
\end{align}
\end{subequations}

Let us now discuss the spectra of electric and magnetic fields that 
arise in the standard scenarios.
Evidently, the spectra will depend on the choice of the coupling 
function~$J(\phi)$.
Often, the coupling function is assumed to be a simple function of the 
scale factor as follows~\cite{Martin:2007ue,Subramanian:2009fu}:
\begin{equation}
J(\eta)= \l[\f{a(\eta)}{a(\ee)}\r]^{\bar{n}},\label{eq:Ja}
\end{equation}
where~$\ee$ is the conformal time at the end of inflation and~$\bar{n}$ is a
number.
Note that the overall coefficient has been chosen so that the coupling
function $J$ reduces to unity at the end of inflation (i.e. at~$\ee$). 
In situations wherein the scale factor can be expressed in the de Sitter 
form, the coupling function is given by 
\begin{equation}
J(\eta)=\l(\f{\eta}{\ee}\r)^{-\bar{n}}.\label{eq:Je}
\end{equation}
For such a choice of the coupling function, the solution to Eq.~\eqref{eq:de-cAk} 
that satisfies the Bunch-Davies initial conditions in the domain wherein $k \gg 
\sqrt{J''/J}$ [which corresponds to the sub-Hubble limit $(-k\,\eta)\ll 1$ for 
the $J$ we are working with] can be expressed as
\begin{equation}
\cA_k(\eta) 
= \sqrt{-\f{\pi\,\eta}{4}}\,
\mathrm{e}^{i\,(\bar{n}+1)\,\pi/2}\,
H^{(1)}_{\nu}(-k\,\eta),\label{eq:nhs}
\end{equation}
where $\nu=\bar{n}+(1/2)$, and $H_\nu^{(1)}(z)$ denotes the Hankel function 
of the first kind.

The spectra of the electromagnetic fields can be evaluated at late times 
when $k \ll \sqrt{J''/J}$, i.e. in the super-Hubble limit wherein $(-k\,\eta) 
\ll 1$.
Upon substituting the mode functions given in Eq.~\eqref{eq:nhs} in the 
expressions in Eqs.~\eqref{eq:pbe} and considering the super-Hubble limit, 
the spectra of the magnetic and electric fields~$\pb(k)$ and $\pe(k)$ can 
be obtained to be~\cite{Martin:2007ue,Subramanian:2009fu}
\begin{subequations}\label{eq:pbe-f}
\begin{eqnarray}
\pb(k) &=& \f{\HI^4}{8\,\pi}\, \mathcal{F}(m)\,(-k\,\ee)^{2\,m+6},\label{eq:pb-f}\\
\pe(k) &=& \f{\HI^4}{8\,\pi}\, \mathcal{G}(m)\,(-k\,\ee)^{2\,m+4}.
\end{eqnarray}
\end{subequations}
The quantities $\mathcal{F}(m)$ and $\mathcal{G}(m)$ are given by
\begin{subequations}\label{eq:f-g}
\begin{eqnarray}
\mathcal{F}(m)
&=&\f{1}{2^{2\,m+1}\,\mathrm{cos}^2(m\,\pi)\,\Gamma^2(m+3/2)},\\
\mathcal{G}(m)
&=&\f{1}{2^{2\,m-1}\,\mathrm{cos}^2(m\,\pi)\,\Gamma^2(m+1/2)},
\end{eqnarray}
\end{subequations}
with 
\begin{equation}
m=\begin{cases}
\bar{n}, & \text{for $\bar{n}<-\frac{1}{2}$},\\
-\bar{n}-1, & \text{for $\bar{n}>-\frac{1}{2}$},
\end{cases}
\end{equation}
in the case of $\pb(k)$, and with 
\begin{equation}
m=\begin{cases}
\bar{n}, & \text{for $\bar{n}<\f{1}{2}$},\\
1-\bar{n}, & \text{for $\bar{n}>\f{1}{2}$},
\end{cases}
\end{equation}
in the case of $\pe(k)$.
Note that the spectral indices for the magnetic and electric fields, say,
$\nb$ and $\ne$, can be written as
\begin{equation}\label{eq:nb}
\nb 
=\begin{cases}
2\,\bar{n}+6, & \text{for $\bar{n}<-\f{1}{2}$},\\
4-2\,\bar{n}, & \text{for $\bar{n}>-\f{1}{2}$},
\end{cases}
\end{equation}
and 
\begin{equation}
\ne
=\begin{cases}
2\,\bar{n}+4, & \text{for $\bar{n}<\f{1}{2}$},\\
6-2\,\bar{n}, & \text{for $\bar{n}>\f{1}{2}$}.
\end{cases}
\end{equation}

There are a few points that we need to clarify regarding the results that we 
have arrived at above.
To begin with, note that, we obtain a scale invariant spectrum for the magnetic
field when $\bar{n}=2$ and when $\bar{n}=-3$.
However, when $\bar{n}=-3$, the spectrum of the electric field behaves as $\pe(k)
\propto (-k\,\ee)^{-2}$, which grows indefinitely at late times.
In other words, the energy density in the electric field becomes large leading to
significant backreaction on the background (in this regard, see, for instance,
Refs.~\cite{Markkanen:2017kmy,Tripathy:2021sfb}).
Therefore, when we require a scale invariant spectrum for the magnetic field,
we shall focus on the $\bar{n}=2$ case. 
Secondly, recall that the original non-conformal coupling function $J$ depends
on~$\phi$ [cf. Eq.~\eqref{eq:ema-1}].
Whereas, our choices for $J$ in Eqs.~\eqref{eq:Ja} and~\eqref{eq:Je} depend on 
the scale factor and conformal time.
Such a behavior for $J(\eta)$ is achievable in most models of inflation that 
permit slow roll~\cite{Kanno:2009ei,Watanabe:2009ct,Martin:2011sn}.
However, there exist situations, in particular, when an inflationary potential
permits a period of ultra slow roll, wherein a behavior such as $J\propto a^2
\propto \eta^{-2}$ may not be achievable (for discussions in this context, 
see Ref.~\cite{Tripathy:2021sfb}).
To circumvent such challenges, either, one has to consider inflation involving 
more than one field (for a discussion on this point, see 
Ref.~\cite{Tripathy:2022iev}) or, as we shall discuss in the following section, 
consider more non-trivial forms for the coupling functions involving the kinetic
energy of the inflaton.

%%%%%%%%%%%%%%%%%%%%%%%%%%%%%%%%%%%%%%%%%%%%%%%%%%%%%%%%%%%%%%%%%%%%%%%%%%%%%%%

\section{Construction of the non-conformal coupling function in pure 
ultra slow roll inflation}\label{sec:JX}

We had pointed out earlier that a phase of ultra slow roll inflation is 
achieved as the inflaton approaches a point of inflection in the potential.
Hence, during the period of ultra slow roll, the velocity of the inflaton
drops rapidly and the field hardly evolves.
In such situations, the non-conformal coupling function $J(\phi)$ becomes
nearly a constant and, as a result, the quantity $J''/J$ turns out to be 
very small.
This behavior effectively restores the conformal invariance of the 
electromagnetic action~\cite{Tripathy:2021sfb}.
One finds that the electromagnetic spectra behave as $k^4$ over wave 
numbers that leave the Hubble radius after the onset of ultra slow 
roll.
Moreover, the amplitude of the magnetic fields on large scales (which
leave the Hubble radius during the initial period of slow roll, prior 
to the transition to the phase of ultra slow roll), is considerably 
suppressed.

To circumvent this difficulty, we shall now construct a non-conformal 
coupling function~$J$ that involves the time derivative of the inflaton.
Such a coupling function proves to be particularly effective in the 
scenario of {\it pure}\/ ultra slow roll inflation that is of interest 
in this work.
Recall that, in ultra slow roll inflation, the first slow roll parameter 
$\epsilon_1$ behaves as in Eq.~\eqref{eq:e1-p}, where $p>3$.
It should then be clear that, if we have a coupling function that
behaves as $\epsilon_1^{n/2}$, for a suitable choice of the index~$n$, 
we will be able to achieve the behavior of, say, $J \propto \eta^{-2}$, 
as is required to lead to a scale invariant spectrum for the magnetic 
field.

Since $\epsilon_1=\dot{\phi}^2/(2\,H^2\,\Mpl^2)$ and the Hubble parameter~$H$
is nearly a constant in ultra slow roll inflation, we can assume that the 
non-conformal coupling function~$J$ depends on the scalar quantity
$X=-(1/2)\,\pa_\sigma\phi\, \pa^\sigma\phi=\dot{\phi}^2/2$.
Therefore, we can rewrite the action governing the electromagnetic field 
in the following form\footnote{We should mention that a different action
involving derivatives of the inflaton has been considered 
earlier~\cite{Tasinato:2014fia}.}:
\begin{equation}
S[A^\mu] = -\f{1}{16\,\pi} \int \d^4x \sqrt{-g}\, J^2(X)\, 
F_{\mu\nu}\,F^{\mu\nu}.\label{eq:ema-2} 
\end{equation}
We shall assume that the coupling function is given by
\begin{equation}
J(X)= J_0\,\l(-\f{\pa_\sigma\phi\,\pa^\sigma\phi}{2}\r)^{n/2},
\label{eq:Ji}
\end{equation}
where~$J_0$ is a constant which we shall choose so that~$J$ reduces 
to unity at the end of inflation.
It should be evident that, since the non-conformal coupling function~$J(X)$ 
depends only on time, the electromagnetic field described by the 
action in Eq.~\eqref{eq:ema-2} can be treated in the same manner as in the 
case of the field described by the earlier action in Eq.~\eqref{eq:ema-1}.
For instance, if we work in the Coulomb gauge, the Fourier modes $\bar{A}_k$ 
of the vector potential $A_i$ will continue to satisfy the same equation of 
motion, viz.~Eq.~\eqref{eq:de-Abk}.
Also, on quantization, we can elevate the vector potential~$A_i$ to the 
operator~$\hat{A}_i$ as in Eq.~\eqref{eq:fAk}.
Note that, in the ultra slow roll inflationary scenarios of interest, 
we can write the coupling function in Eq.~\eqref{eq:Ji} as
\begin{equation}
J(\eta)=J_0\,\l[\HI^2\, \Mpl^2\, \epsilon_1(\eta)\r]^{n/2},\label{eq:Jusr}
\end{equation}
and we shall choose $J_0=\l[\HI^2\, \Mpl^2\, \epsilon_{1}(\ee)\r]^{-n/2}$
so that $J(\ee)=1 $.  
Therefore, the coupling function can be simply written as follows:
\begin{equation}
J(\eta) = \l[\f{\epsilon_1(\eta)}{\epsilon_{1}(\ee)}\r]^{n/2}.\label{eq:Jf}
\end{equation}

On varying the action in Eq.~\eqref{eq:ema-2} with respect to the metric tensor, 
we obtain the stress-energy tensor for the electromagnetic field to be
\begin{align}
T_{\alpha\beta}&=-\f{J_0^2}{16\,\pi}\,g_{\alpha_\beta}\,
\l(-\pa_\sigma\phi\,\pa^\sigma\phi\r)^n\,F_{\mu\nu}\,F^{\mu\nu}\nn\\
&\quad -\f{J_0^2\,n}{8\,\pi}\,\l(-\pa_\sigma\phi\,\pa^\sigma\phi\r)^{n-1}\,
\l(\pa_\alpha\phi\,\pa_\beta\phi\r)\,F_{\mu\nu}\,F^{\mu\nu}\nn\\
&\quad +\f{J_0^2}{4\,\pi}\l(\pa_\sigma\phi\,\pa^\sigma\phi\r)^n
g^{\mu\nu}\,F_{\alpha\mu}\,F_{\beta\nu}.\label{eq:set}
\end{align}
We find that, in the background described by the FLRW 
line-element described by Eq.~\eqref{eq:flrw-le}, 
the energy density of the electromagnetic field, viz. $\rem = -T^t_t$, 
can be expressed as
\begin{equation}
\rem =(1+2\, n)\,\re + (1-2\, n)\,\rb,\label{eq:rem-2}
\end{equation}
where~$\rb$ and~$\re$ are the energy densities we had introduced earlier 
in Eqs.~\eqref{eq:rbe}.
Since the factors $(1-2\, n)$ and $(1+2\, n)$ in the above expression are 
numbers of the order of unity, we shall continue to define the power spectra 
$\pb(k)$ and $\pe(k)$ of the magnetic and electric fields as we had done in 
Eq.~\eqref{eq:pbe-d}. 
Clearly, in such a case, the power spectra $\pb(k)$ and $\pe(k)$ will be 
given by Eqs.~\eqref{eq:pbe}, with $J(\eta)$ given by Eq.~\eqref{eq:Jf}.
We shall evaluate the power spectra at the end of inflation (i.e. at $\ee$). 
As~$n$ can be positive or negative, a concern may arise if the expectation
value of the total energy density---i.e.~$\langle \hat{\rho}_{_\mathrm{{EM}}}
\rangle$, with $\rem$ given by Eq.~\eqref{eq:rem-2}---is positive definite.
We shall discuss this issue in App.~\ref{app:rho}.
We shall show that, in all the scenarios we consider, the quantity $\langle 
\hat{\rho}_{_\mathrm{{EM}}}\rangle$ always remains positive.

Recall that, in ultra slow roll inflation, the first slow roll parameter
behaves as $\epsilon_1 \propto \eta^p$ [cf. Eq.~\eqref{eq:e1-p}]. 
Also, as we discussed in Sec.~\ref{subsec:PBPE}, in order to obtain a 
scale invariant spectrum for the magnetic field, we need the coupling 
function to behave as $J \propto \eta^{-2}$. 
It should then be clear from Eq.~\eqref{eq:Jf} that we can obtain a scale 
invariant spectrum for the magnetic field if we choose $n\,p=-4$. 
In such a case, we find that the mode function~$\bar{A}_k$ and the scaled
variable~$\cA_k$ are given by
\begin{subequations}
\begin{align}
\cA_k(\eta)
&=-\f{1}{\sqrt{2\, k^5}\,\eta^2}\, 
\l(3 + 3\, i\, k\, \eta - k^2\, \eta^2\r)\, \mathrm{e}^{-i\, k\, \eta},\\
\bar{A}_k(\eta) 
&=-\f{1}{\sqrt{2\, k^5}\,\ee^2}\,
\l(3 + 3\, i\, k\, \eta - k^2\, \eta^2\r) \mathrm{e}^{-i\, k\, \eta}.
\label{eq:Abk}
\end{align}   
\end{subequations}
On using these solutions in Eqs.~\eqref{eq:pbe}, we obtain the power
spectra of the magnetic and electric fields to be
\begin{equation}
\pb(k) \simeq \f{9\,\HI^4}{4\,\pi^2},\;
\pe(k) \simeq \f{\HI^4}{4\,\pi^2}\,(-k\,\ee)^{2}.\label{eq:pbe-nm4}
\end{equation}
Note that, while the power spectrum of the magnetic field $\pb(k)$ is scale 
invariant, the spectrum of the electric field behaves as $\pe(k)\propto k^2$.
However, since $\ee$ is small, on large scales, the strength of~$\pe(k)$ is 
significantly suppressed when compared to that of~$\pb(k)$.

%%%%%%%%%%%%%%%%%%%%%%%%%%%%%%%%%%%%%%%%%%%%%%%%%%%%%%%%%%%%%%%%%%%%%%%%%%%%%%%

\section{Cubic order action and the cross-correlation between curvature
perturbations and magnetic fields}\label{sec:RBB}

It is well known that, if the primordial perturbations were Gaussian, all
their statistical properties would be contained in the two-point functions. 
In such a case, while the higher-order even-point functions can be 
expressed in terms of the two-point functions, the higher-order 
odd-point functions would vanish. 
However, if the perturbations were not Gaussian, the odd-point functions 
can, in general, be expected to be non-zero. 
This carries importance because, in many situations, the non-Gaussianities 
can significantly alter the predictions of the cosmological observables 
(in this context, see, for example, Refs.~\cite{Taruya:2008pg,Unal:2018yaa,
Adshead:2021hnm,Ragavendra:2021qdu,Taoso:2021uvl,Ferrante:2022mui,
Yamauchi:2022fri,Franciolini:2023pbf,Li:2023xtl,Pi:2024jwt,Perna:2024ehx}).
Evidently, these arguments also apply to primordial magnetic fields. 
At the level of the two-point functions, one finds that many models of 
inflation generate magnetic fields with scale invariant spectra. 
So, to distinguish between the various models from the perspective of 
electromagnetic fields, it is not adequate to study only the two-point 
functions. 
We need to evaluate the higher-order correlations, such as the three-point 
functions. 
The three-point cross-correlation between the scalar perturbations and the 
magnetic fields has been studied extensively in the case of slow roll 
inflation~\cite{Seery:2008ms,Caldwell:2011ra,Motta:2012rn,Jain:2012vm,
Jain:2012ga,Kunze:2013hy,Chowdhury:2018mhj,Sai:2023vyf}.
Our aim in this work will be to evaluate the three-point cross-correlation 
in the scenario of {\it pure}\/ ultra slow roll inflation. 
We shall focus on particular cases that exhibit certain behaviour for the 
power spectra of the scalar perturbations and the electromagnetic fields. 
Specifically, we shall investigate the behavior of the cross-correlation in 
the squeezed limit wherein the wave number of the scalar perturbation is much
smaller than the wave numbers of the two electromagnetic modes.

%%%%%%%%%%%%%%%%%%%%%%%%%%%%%%%%%%%%%%%%%%%%%%%%%%%%%%%%%%%%%%%%%%%%%%%%%%%%%%%

\subsection{Action at the cubic order}

As is often done, we shall compute the three-point function using the standard
methods of perturbative quantum field theory (for calculations involving scalars
and tensors, see, for example, Refs.~\cite{Maldacena:2002vr,Seery:2005wm,
Chen:2005fe,Chen:2006nt,Langlois:2008wt}, and for discussions involving magnetic 
fields, see Refs.~\cite{Caldwell:2011ra,Motta:2012rn,Jain:2012vm,Jain:2012ga,
Kunze:2013hy,Chowdhury:2018mhj,Chowdhury:2018blx,Sai:2023vyf}).
Our first task is to obtain the interaction Hamiltonian at the cubic order 
that governs the cross-correlation of interest.
We shall start with the action in Eq.~\eqref{eq:ema-2}, take into account the 
scalar perturbations in the spatially flat, FLRW line-element, and arrive at the 
cubic order action involving the curvature perturbations and the electromagnetic 
fields.  
We shall make use of the Arnowitt–Deser–Misner~(ADM) formalism to obtain the 
action~\cite{Arnowitt:1960es}.
Recall that, in the ADM formalism, the line-element associated with a generic 
spacetime is described in terms of the lapse function $N$, the shift vector 
$N^i$ and the spatial metric~$g_{ij}$ as follows:
\begin{equation}
\d s^2= -N^2\,\l(\d x^0\r)^2+g_{ij}\,(N^i\,{\d x^0+\d x^i})\,
(N^j\,\d x^0+\d x^j).
\end{equation}
Note that the metric elements $g_{00}$ and $g_{0i}$ associated with this
line-element are given by 
\begin{equation}
g_{00} = -N^2+g_{ij}\,N^i\,N^j,\quad g_{0i} = g_{ij}N^j.
\end{equation}
We shall work in the gauge wherein the perturbations in the inflaton are absent.
In such a case, to account for the scalar perturbations, we can write the spatial 
element $g_{ij}$ of the spatially flat, FLRW line-element as follows:
\begin{equation}
g_{ij} =\delta_{ij}\,a^2\,\mathrm{e}^{2\,\mathcal{R}},
\end{equation}
where, as we have mentioned, $\cR$ denotes the curvature perturbation.
The lapse and the shift functions~$N$ and~$N_i$ can be determined from the 
constraint equations at the first order. 
If we work with the conformal time coordinate and write $N = a\,(1+N_1)$ and 
$g_{ij}\,N^j= a^2\, \partial_i\chi$, where $N_1$ and $\chi$ are the inhomogeneous 
terms, we find that the constraint equations determine $N_1$ and $\chi$ to be
\begin{equation}
N_1 = \f{\cR'}{a\,H},\; 
\chi= -\f{\cR}{a\,H}+\epsilon_1\,\nabla^{-2} \cR'.
\end{equation}

\begin{widetext}
On substituting the above components of the metric tensor in the 
action in Eq.~\eqref{eq:ema-2}, we find that the action {\it up to}\/ 
the cubic order involving the curvature perturbation~$\cR$ and 
the electromagnetic vector potential~$A_i$ can be expressed as
\begin{align}
S[\cR,A_i] &= -\f{1}{16\,\pi}\int \d\eta\, \int \d^3 {\bm x}\, 
J^2(\eta)\,
\bigg\{-2\,\l[1+\cR - (1+2\,n)\,\frac{\mathcal{R}'}{a\,H}\r]\,
\delta^{ij}\,A_i'\,A_j'\nn\\
&\quad+ 4\,\delta^{il}\,\delta^{jm} \pa_i\chi\,A_j'\,F_{lm}
+ \l[1-\mathcal{R}+(1-2\,n)\,\f{\mathcal{R'}}{a\,H}\r]\,
\delta^{im}\,\delta^{jn}\,F_{ij}\,F_{mn}\biggr\},
\end{align}
where, in the cases of our interest, the non-conformal coupling 
function~$J(\eta)$ is given by Eq.~\eqref{eq:Jf}.
To calculate the three-point cross-correlation, we shall require the 
cubic order Hamiltonian associated with the above action.
From the action, we obtain the momentum conjugate to the vector 
potential~$A_i$, say, $\pi_i$, to be
\begin{equation}
\pi_i =-\f{a^2(\eta)}{4\,\pi}\, J^2(\eta)\,
\l\{-\l[1+\cR-(1+2\,n)\,\f{\mathcal{R}'}{a\,H}\r]\,{A_i'}
+\l[1+\cR-(1+2\,n)\,\f{\cR'}{a\,H}\r]\, \delta^{jl}\,\pa_j {\chi} F_{li}\r\}.
\end{equation}
We can make use of this conjugate momentum to construct the Hamiltonian and 
identify the cubic order terms as the interaction Hamiltonian~$H_\mathrm{int}$. 
We obtain the interaction Hamiltonian to be 
\begin{align}
H_\mathrm{int} 
&= -\f{1}{16\,\pi}\, \int \d^3{\bm x}\, J^2(\eta)\,
\biggl\{2\,\l[\cR-(1+2\,n)\,\f{\cR'}{a\,H}\r]\,\delta^{ij}\,A_i'\,A_j'
-4\,\delta^{il}\,\delta^{jm} \pa_i\chi\,A_j'\,F_{lm}\nn\\ 
&\quad+\l[\cR-(1-2\,n)\,\f{\cR'}{a\,H}\r]\,
\delta^{im}\,\delta^{jn}\,F_{ij}\,F_{mn}\biggr\}.\label{eq:Hint}
\end{align} 
There are two clarifying remarks that we wish to make regarding the 
interaction Hamiltonian~$H_\mathrm{int}$ we have obtained.
Firstly, note that, in the above Hamiltonian, if we ignore the terms 
containing~$n$ within the curly brackets, it reduces to the standard 
form associated with the original action in Eq.~\eqref{eq:ema-1} 
involving~$J(\phi)$ (in this regard, see, for example, 
Refs.~\cite{Motta:2012rn,Jain:2012vm}).
Secondly, as with any action at the cubic order, the interaction 
Hamiltonian proves to be negative of the corresponding cubic order 
Lagrangian.

%%%%%%%%%%%%%%%%%%%%%%%%%%%%%%%%%%%%%%%%%%%%%%%%%%%%%%%%%%%%%%%%%%%%%%%%%%%%%%%

\subsection{Three-point cross-correlation}

We shall now turn to the computation of the three-point cross-correlation using 
the interaction Hamiltonian in Eq.~\eqref{eq:Hint}. 
In real space, the cross-correlation between the curvature perturbation and 
magnetic fields is defined as
\begin{equation}
\langle \hat{\cR}(\eta,\vx)\, \hat{B}^i(\eta,\vx)\,\hat{ B}_i(\eta,\vx)\rangle 
= \int \f{\d^3{\bm k}_1}{(2\,\pi)^{3/2}}\int \f{\d^3{\bm k}_2}{(2\,\pi)^{3/2}}
\int \f{\d^3{\bm k}_3}{(2\,\pi)^{3/2}}
\langle \hat{\cR}_{\vk_1}(\eta)\, \hat{B}^i_{\vk_2}(\eta)\,
\hat{B}_{i\,{\vk_3}}(\eta) \rangle\,
\mathrm{e}^{i\, (\vk_1+\vk_2+\vk_3) \cdot \vx},
\end{equation}
where the components $B_i$ of the magnetic fields are related to those of the 
vector potential $A_i$ through the relation 
\begin{equation}
B_i = \f{1}{a}\, \epsilon_{ijl}\, \pa_j A_l.
\end{equation}
We had mentioned above that we shall make use of perturbative quantum field 
theory to arrive at the three-point cross-correlation.
According to quantum field theory, given the cubic order Hamiltonian $H_\mathrm{int}$, 
the three-point cross-correlation in Fourier space, evaluated at the conformal time~$\ee$
corresponding to the end of inflation, is given by
\begin{equation}
\langle \hat{\cR}_{\vk_1}(\ee)\, \hat{B}^i_{\vk_2}(\ee)\,
\hat{B}_{i\,{\vk_3}}(\ee) \rangle\,
= -i \int_{\ei}^{\ee} \d\eta\, 
\langle[\hat{\cR}_{\vk_1}(\ee)\, \hat{B}^i_{\vk_2}(\ee)\,
\hat{B}_{i\,{\vk_3}}(\ee),\hat{H}_\mathrm{int}(\eta)]\rangle,\label{eq:3ptint}
\end{equation}
where $\hat{H}_{\mathrm{int}}$ is the operator associated with the interaction 
Hamiltonian in Eq.~\eqref{eq:Hint}, and $\ei$ denotes the time when the initial 
conditions are imposed on the perturbations.
Moreover, the square brackets denote the commutator between the operators denoting
modes of the curvature perturbation, the magnetic field, and the Hamiltonian operator.

To describe the cross-correlation between the modes of the curvature
perturbation and the magnetic field, we shall introduce the 
function~$\cB(\vk_1,\vk_2,\vk_3)$ that is defined through the 
relation\footnote{In the definition in Eq.~\eqref{eq:3ptB} of the three-point function, 
we have introduced an additional factor of $(4\,\pi)$ in comparison to the earlier 
works in the literature (in this context, see Refs.~\cite{Motta:2012rn,Caldwell:2011ra,
Jain:2012vm,Jain:2012ga,Nandi:2021lpf}).
Note that our electromagnetic actions in Eqs.~\eqref{eq:ema-1} and~\eqref{eq:ema-2}) 
contain an additional factor of $1/(4\,\pi)$ in comparison to the action in 
the previous works. 
In order to ensure that the expression for the power spectrum of the magnetic
field remains the same, we had included an overall factor of $\sqrt{4\,\pi}$ 
in the expression for the vector potential~$A_i$ in Eq.~\eqref{eq:fAk}.
However, such a definition for $A_i$ introduces another factor of $(4\,\pi)$ when
we evaluate the quantity on the right hand side of Eq.~\eqref{eq:3ptint}, which 
is not present if we instead follow the normalization conventions used in the 
aforementioned works.
Therefore, in order to ensure that the quantity $\cB(\vk_1,\vk_2,\vk_3)$ remains
the same across the different conventions, we have introduced the factor of 
$(4\,\pi)$ in Eq.~\eqref{eq:3ptB}.}
\begin{equation}
\langle \hat{\cR}_{\vk_1}(\ee)\, \hat{B}^i_{\vk_2}(\ee)\,
\hat{B}_{i\,{\vk_3}}(\ee) \rangle\,
=\f{4\,\pi}{(2\,\pi)^{{3}/{2}}}\,\cB(\bm{k}_1,\bm{k}_2,\bm{k}_3)\,
\delta^{(3)}(\vk_1+\vk_2+\vk_3).\label{eq:3ptB}
\end{equation}
On making use of the expression in Eq.~\eqref{eq:3ptint}, the interaction 
Hamiltonian in Eq.~\eqref{eq:Hint}, the definition of the three-point
function in Eq.~\eqref{eq:3ptB} and, finally, Wick's theorem, we find that 
the function~$\cB(\vk_1,\vk_2,\vk_3)$ 
can be written as follows~\cite{Motta:2012rn,Caldwell:2011ra,Jain:2012vm,
Jain:2012ga,Chowdhury:2018blx,Nandi:2021lpf}:
\begin{align}
\cB(\vk_1,\vk_2,\vk_3)
= \sum_{C=1}^{6} \cB_{C}(\vk_1,\vk_2,\vk_3)
= \f{1}{a^4(\ee)}\, \sum_{C=1}^{6} \l[f_{k_1}(\ee)\,\bar{A}_{k_2}(\ee)\,
\bar{A}_{k_3}(\ee)\, \cG_{C}(\vk_1,\vk_2,\vk_3) 
+ \mbox{complex conjugate}\r],\label{eq:cB} 
\end{align}
where the quantities $\mathcal{G}_{C}(\vk_1,\vk_2,\vk_3)$, with $C=\{1,6\}$, 
are described by the integrals
\begin{subequations}\label{eq:cG}
\begin{align}
\cG_1(\vk_1,\vk_2,\vk_3)
&= -2\,i\, (\bm{k}_2\cdot\bm{k}_3)\,
\int_{\ei}^{\ee} \d\eta\, J^2(\eta)\,
f_{k_1}^\ast(\eta)\,\bar{A}_{k_2}'^{\ast}(\eta)\,
\bar{A}_{k_3}'^{\ast}(\eta),\\
\cG_2(\bm{k}_1,\bm{k}_2,\bm{k}_3)
&= 2\,i\,(1+2\,n)\,(\bm{k}_2\cdot\bm{k}_3)\,
\int_{\ei}^{\ee} \f{\d\eta}{a\,H}\,  J^2(\eta)\,
f_{k_1}'^{\ast}(\eta)\,\bar{A}_{k_2}'^{\ast}(\eta)\,
\bar{A}_{k_3}'^{\ast}(\eta),\\
\cG_3(\vk_1,\vk_2,\vk_3)
&= i\,\l[(\bm{k}_1\cdot\bm{k}_2)\,k_3^2
+(\bm{k}_1\cdot\bm{k}_3)\,(\bm{k}_2\cdot\bm{k}_3)\r]\,
\int_{\ei}^{\ee} \f{\d\eta}{a\,H}\, J^2(\eta)\,
f_{k_1}^{\ast}(\eta)\,\bar{A}_{k_2}'^{\ast}(\eta)\,
\bar{A}_{k_3}^{\ast}(\eta)\nn\\ 
&\quad+\vk_2 \leftrightarrow \vk_3,\\
\cG_4(\vk_1,\vk_2,\vk_3)
&= \f{i}{k_1^2}\,\l[(\bm{k}_1\cdot\bm{k}_2)\,k_3^2
+(\bm{k}_1\cdot\bm{k}_3)\,(\bm{k}_2\cdot\bm{k}_3)\r]\,
\int_{\ei}^{\ee} \d\eta\,J^2(\eta)\,
\epsilon_1(\eta)\, f_{k_1}'^{\ast}(\eta)\,\bar{A}_{k_2}'^{\ast}(\eta)\,
\bar{A}_{k_3}^{\ast}(\eta)\nn\\ 
&\quad+\vk_2 \leftrightarrow \vk_3,\\
\cG_5(\vk_1,\vk_2,\vk_3)
&= i\,\l[k_2^2\,k_3^2 + (\bm{k}_2\cdot\bm{k}_3)^2\r]\,
\int_{\eta_\mathrm{i}}^{\ee} \d\eta\,  J^2(\eta)\,
f_{k_1}^\ast(\eta)\,\bar{A}_{k_2}^\ast(\eta)\,\bar{A}_{k_3}^\ast(\eta),\\
\cG_6(\vk_1,\vk_2,\vk_3)
&= -i\,(1-2\,n)\, \l[k_2^2\,k_3^2+(\bm{k}_2\cdot\bm{k}_3)^2\r]\,
\int_{\ei}^{\ee} \f{\d\eta}{a\,H}\, J^2(\eta)\,
f_{k_1}'^{\ast}(\eta)\,\bar{A}_{k_2}^{\ast}(\eta)\,
\bar{A}_{k_3}^{\ast}(\eta).
\end{align}
\end{subequations} 
We should mention that these six integrals are exact expressions and they
have been arrived at without making any approximations. 

Using the solutions for~$f_k$ and~$\bar{A}_k$ in the pure ultra slow roll 
scenarios that we discussed in the previous two sections, we can compute 
the integrals~$\cG_C(\vk_1,\vk_2,\vk_3)$. 
In App.~\ref{app:bnl-calG} we have presented the explicit forms 
of~$\cG_C(\vk_1,\vk_2,\vk_3)$ for two cases of~$p$ and suitable choices 
for~$n$ that lead to scale invariant spectra of the magnetic fields. 
Using the expressions for~$\cG_C(\vk_1,\vk_2,\vk_3)$, we can compute the 
different contributions to the cross-correlation~$\cB(\vk_1,\vk_2,\vk_3)$.
We shall discuss these results in the next section. 
\end{widetext}

%%%%%%%%%%%%%%%%%%%%%%%%%%%%%%%%%%%%%%%%%%%%%%%%%%%%%%%%%%%%%%%%%%%%%%%%%%%%%%%

\subsection{Non-Gaussianity parameter} %$\bnl(\vka,\vkb,\vkc)$}

In addition to the cross-correlation~$\cB(\vk_1,\vk_2,\vk_3)$, we shall compute 
the associated non-Gaussianity parameter~$\bnl(\vk_1,\vk_2,\vk_3)$.
The parameter is usually expressed as a dimensionless quantity involving the 
ratio of the three-point cross-correlation and the scalar power spectrum and 
the power spectrum of the magnetic field~\cite{Caldwell:2011ra,Motta:2012rn,
Jain:2012ga,Jain:2012vm,Chowdhury:2018blx}. 
The parameter reflects the amplitude and shape of the three-point function
and it helps in distinguishing among various models of inflation as well
as between inflationary and alternative scenarios of the early universe
which lead to the same power spectra.

To arrive at the form of the non-Gaussianity parameter~$\bnl(\vk_1,\vk_2,\vk_3)$, 
let us express the magnetic field in terms of the Gaussian components and the 
parameter as follows:
\begin{align}
\hat{B}_{i\,\vk}(\eta) 
& = \hat{B}_{i\,\vk}^{_{\mathrm{G}}}(\eta)
+\f{1}{(2\,\pi)^{3/2}}\int \d^3\bm{p}\, \hat{\cR}_{\bm p}^{_{\mathrm{G}}}(\eta)\,
\hat{B}_{i\,{\bm{k}-\bm{p}}}^{_{\rm G}}(\eta)\nn\\
&\quad\times\,\bnl({\bm p},\vk, \vk-{\bm p}),\label{eq:bnl-def}
\end{align}
where~$\hat{B}^{_{\mathrm{G}}}_{i\,\vk}$ and $\hat{\mathcal{R}}_{\vk}^{_{\mathrm{G}}}$ 
represent the Gaussian parts of the Fourier modes of the magnetic field and the 
curvature perturbation, respectively. 
If we compute the cross-correlation~$\cB(\vk_1,\vk_2,\vk_3)$ using the 
above expression and utilize Wick's theorem which applies to the Gaussian
components, we obtain the expression for~$\bnl(\vk_1,\vk_2,\vk_3)$ to be
\begin{align}
\bnl(\vk_1,\vk_2,\vk_3) 
& = \f{1}{8\,\pi^4}\, J^2(\eta)\nn\\
&\quad\times\f{k_1^3\,k_2^3\,k_3^3\,\cB(\bm{k}_1,\bm{k}_2,\bm{k}_3)}
{\ps(k_1)\,\l[k_3^3\,\pb(k_2)+k_2^3\,\pb(k_3)\r]},\label{eq:bnl}
\end{align}
where, recall that, $\ps(k)$ and $\pb(k)$ represent the scalar power spectrum 
and the power spectrum of the magnetic field.

An important property of the three-point functions is their behavior in
the so-called squeezed limit wherein one of the three wavelengths is much
longer than the other two.
In the squeezed limit, under certain conditions, it has been shown that the
three-point functions can be expressed entirely in terms of the two-point 
functions. 
This is often referred to as the consistency relation~\cite{Starobinsky:1994bd,
Maldacena:2002vr,Creminelli:2004yq,Tsamis:2008it,Seery:2008ax,Giddings:2010nc,
Giddings:2011zd,Creminelli:2012ed,Kehagias:2012pd,Jain:2012vm,Jain:2012ga,
Sreenath:2014nka,Sreenath:2014nca,Chowdhury:2018blx,Sai:2023vyf}. 
Such a consistency condition primarily arises because of the reason that, in the 
scenarios involving slow roll, the amplitude of the long-wavelength perturbation
freezes in the super-Hubble regime.
As far as the three-point cross-correlation of our interest is concerned, such
a consistency relation can be expected to arise when one considers the mode
associated with the curvature perturbation to be the long wavelength mode. 
Indeed, such a relation has been established in slow roll
inflation (in this context, see Refs.~\cite{Jain:2012vm,Jain:2012ga}).
In the following sections, apart from computing the three-point cross-correlation 
and the associated non-Gaussianity parameter, we shall also explicitly examine if
the standard consistency condition is valid in the scenarios involving ultra slow 
roll inflation.

%%%%%%%%%%%%%%%%%%%%%%%%%%%%%%%%%%%%%%%%%%%%%%%%%%%%%%%%%%%%%%%%%%%%%%%%%%%%%%%

\section{Calculation of the cross-correlation and associated non-Gaussianity
parameter}{\label{sec:calB}}

We shall now evaluate the three-point cross-correlation~$\cB(\vk_1,\vk_2,\vk_3)$ 
and the associated non-Gaussianity parameter~$\bnl(\vk_1,\vk_2,\vk_3)$ in the 
ultra slow roll inflationary scenarios we considered earlier. 
Before discussing the amplitude and shape of the non-Gaussianity parameter, 
we shall clarify a few points regarding the scalar power spectra and the
power spectra of the magnetic field, and arrive at suitable values of the
parameters involved.

%%%%%%%%%%%%%%%%%%%%%%%%%%%%%%%%%%%%%%%%%%%%%%%%%%%%%%%%%%%%%%%%%%%%%%%%%%%%%%%

\subsection{Power spectra}

We shall begin by fixing the Hubble parameter during inflation~$\HI$ 
using the condition that the scalar power at the pivot scale of 
$k_\ast=0.05\,\mpcinv$ is normalized by the CMB observations to be
$\ps(k_\ast) \simeq 2 \times 10^{-9}$.
Fixing~$\HI$ in such a manner is not straightforward and is dependent
on the ultra slow scenario that we are considering.
Recall that, in the ultra slow roll scenarios with $p=6$ and $p=4$,
the scalar power spectrum is given by [cf. Eqs.~\eqref{eq:ps-p6} 
and~\eqref{eq:ps-p4}]
\begin{equation}
\ps(k) \simeq \begin{cases}
\f{\HI^2}{8\,\pi^2\,\eai\,\Mpl^2}\,\l(\f{\ei}{\ee}\r)^6, & \text{for $p=6$},\\
\f{\HI^2}{8\,\pi^2\,\eai\,\Mpl^2}\,\l(\f{\ei}{\ee}\r)^2\,
(k\,\ei)^2, & \text{for $p=4$}.
\end{cases}
\end{equation}
Demanding $\ps(k_\ast) \simeq 2 \times 10^{-9}$ implies that the Hubble 
parameter is to be set to 
\begin{equation}
\HI \simeq \begin{cases}
4\, \pi\, 10^{-9/2}\, \sqrt{\eai}\,
\l(\f{\ee}{\ei}\r)^3\Mpl, & \text{for $p=6$},\\
4\, \pi\, 10^{-9/2}\, \sqrt{\eai}\,\l(\f{\ee}{\ei}\r)\, 
\f{1}{\vert k_\ast\,\ei \vert}\, \Mpl, & \text{for $p=4$}.
\end{cases}
\end{equation}

Let us then turn to the power spectrum of the tensor perturbations. 
Recall that the equation of motion that governs the tensor perturbations
is solely dictated by the behavior of the scale factor~$a(\eta)$.
Also, note that the scale factor during ultra slow roll inflation can
be approximated well by the de Sitter form as $a(\eta) = -1/(\HI\,\eta)$.
Hence, the spectrum of tensor perturbations in the ultra slow roll 
scenarios of interest (for $p=6$ as well as~$4$) is given by the 
standard form in slow roll inflation, viz.  
\begin{equation}
\pt(k) \simeq \f{2\,\HI^2}{\pi^2\,\Mpl^2}.
\end{equation}

Next, we set $\ei = -10^6\,\mathrm{Mpc}$ so that the largest observable
scale of $k \simeq 10^{-4}\,\mpcinv$ is sufficiently deep inside the
Hubble radius at the beginning of inflation.
We shall assume that inflation lasts for about $65$ e-folds.
In such a case, $\ee = -10^{-19}\, \mathrm{Mpc}$, which determines the
smallest scale that exits the Hubble radius during inflation to be $k 
\simeq 10^{19}\, \mpcinv$.
Lastly, we choose $\eai = 10^{-3}$ as an indicative value.
It is related to the tensor-to-scalar ratio at the pivot scale of~$k_\ast$,
which in these cases turns out to be
\begin{equation}
r = \begin{cases}
16\,\eai\,\l(\f{\ee}{\ei}\r)^6, &\text{for $p=6$},\\
16\,\eai\,\l(\f{\ee}{\ei}\r)^2\,
\l(\f{1}{k_\ast\ei}\r)^{2}, &\text{for $p=4$}.
\end{cases}
\end{equation}

We should caution the readers about a few points concerning the
behavior of these spectra and the related choices for the parameters. 
The first issue is regarding the shape of the spectra. 
The case of $p=4$ leads to a scalar power spectrum with a strong 
scale-dependence of~$\ps(k) \propto k^2$ and it is clearly at odds 
with the CMB data. 
So, this case is being examined solely for gaining a theoretical
understanding of the effects of ultra slow roll inflation.
The case of $p=6$ performs better in terms of the shape of the
scalar power spectrum. 
In this case, the scalar power spectrum $\ps(k)$ is strictly scale
invariant.
Although, such a spectrum is disfavored by the CMB data, the value
of~$p$ (or, equivalently, the second slow roll parameter~$\epsilon_2$) 
can be tuned slightly to achieve the required spectral index of $\ns 
\simeq 0.965$.

Another outcome of these choices is that the Hubble parameter~$\HI$
and, hence, the tensor-to-scalar ratio~$r$ are highly suppressed (in 
both the cases of $p=6$ and~$4$) by factors of the ratio~$\ee/\ei$. 
Note that~$\ee/\ei \simeq 10^{-25}$. 
Therefore, for instance, when $p=6$, we have $\HI/\Mpl\simeq 
10^{-80}$, which corresponds to an energy scale that is smaller
than the scale of big bang nucleosynthesis!
So, with any value of $\eai$ less than unity, we have a highly 
suppressed tensor power and hence~$r$ in the ultra slow roll
scenarios.

This brings us to yet another issue that we encounter.
Recall that the amplitude of the scale invariant spectrum of the 
magnetic field is given by $\pb(k) \simeq 9\,\HI^4/(4\,\pi^2)$ [cf.
Eq.~\eqref{eq:pbe-nm4}].
Since $\HI$ is very small, the strengths of the generated magnetic 
fields is proportionately suppressed in ultra slow roll inflation. 
If we focus specifically on the case of $p=6$, we find that the
spectrum of the magnetic field can be expressed as 
\begin{equation}
\pb(k)  \simeq  \f{9\, \HI^4}{4\,\pi^2} 
\simeq 9\,(4\,\pi\,\eai)^2\, \l(\f{\ee}{\ei}\r)^{12}\,
\ps(k_\ast)\,\Mpl^4.
\end{equation}
In contrast, the typical value of $\pb(k)$ obtained in slow roll
inflation with a coupling function behaving as $J(\eta) \propto 
\eta^{-2}$ would be
\begin{eqnarray}
\pb(k)  \simeq  \f{9\,\HI^4}{4\,\pi^2}
 \simeq  9\,(4\,\pi\,\eai)^2\,\ps(k_\ast)\,\Mpl^4.
\end{eqnarray}
Clearly, the amplitude of $\pb(k)$ in ultra slow roll inflation is
suppressed by a factor of $(\ee/\ei)^{12} \sim 10^{-300}$. 
Such a small value of $\pb(k)$ will not be sufficient to serve as
seeds for magnetic fields with the strengths observed in the 
inter-galactic medium today.
Despite these shortcomings of the scenario of pure ultra slow roll, we
pursue the analysis with the aim of eventually utilizing the coupling
function that we have constructed in a realistic model of inflation with
a brief epoch of ultra slow roll.
We shall now proceed with these values of the parameters to calculate and
examine the amplitude and shape of the non-Gaussianity
parameter~$\bnl(\vk_1,\vk_2,\vk_3)$.

%%%%%%%%%%%%%%%%%%%%%%%%%%%%%%%%%%%%%%%%%%%%%%%%%%%%%%%%%%%%%%%%%%%%%%%%%%%%%%%

\subsection{Amplitude and shape of the non-Gaussianity parameter}

We shall evaluate the cross-correlation~$\cB(\vk_1,\vk_2,\vk_3)$ between
the curvature perturbation and the magnetic fields in ultra slow roll 
inflationary scenarios with~$p=6$ and~$4$.
We shall consider $J(\eta)$ as given by Eq.~\eqref{eq:Jf} and focus on
the case of $n\,p=-4$ which leads to a scale invariant spectrum for the
magnetic field.
It should be clear from Eqs.~\eqref{eq:cB} and~\eqref{eq:cG} that the
calculation of the three-point cross-correlation first requires the
evaluation of the integrals~$\cG_C(\vk_1,\vk_2,\vk_3)$.
With the background quantities and the Fourier mode functions~$f_k$ as well
as~$\bar{A}_k$ of the curvature perturbation and the electromagnetic
vector potential at hand, we can carry out the 
integrals~$\cG_C(\vk_1,\vk_2,\vk_3)$.
The integrals are easy to compute.
But, since the expressions describing the integrals prove to be lengthy,
we have listed the results for these integrals in the different cases in
App.~\ref{app:bnl-calG}. 
In what follows, we shall list the results for the various 
contributions to~${\cal B}(\vk_1,\vk_2,\vk_3)$ and discuss the structure
of the corresponding non-Gaussianity parameter~$\bnl(\vk_1,\vk_2,\vk_3)$. 

%%%%%%%%%%%%%%%%%%%%%%%%%%%%%%%%%%%%%%%%%%%%%%%%%%%%%%%%%%%%%%%%%%%%%%%%%%%%%%%

\begin{widetext}

\subsubsection{Case of $p=6$}

When~$p=6$ and~$n=-2/3$ so that $n\,p=-4$, we can use the forms
of the slow roll parameter~$\epsilon_1(\eta)$ and the coupling 
function~$J(\eta)$ [cf. Eqs.~\eqref{eq:e1-p} and~\eqref{eq:Jf}] 
as well as the corresponding Fourier mode functions~$f_k(\eta)$ 
and~$\bar{A}_k(\eta)$ describing the curvature perturbation and
the electromagnetic vector potential [cf. Eqs.~\eqref{eq:fk-p6}
and~\eqref{eq:Abk}] to evaluate the integrals~$\cG_C(\vk_1,\vk_2,\vk_3)$. 
Once we have evaluated the quantities~$\cG_C(\vk_1,\vk_2,\vk_3)$,
we can substitute them and the mode functions~$f_k(\eta)$  
and~$\bar{A}_k(\eta)$ in Eq.~\eqref{eq:cB} to arrive at the 
different contributions~$\cB_C(\vk_1,\vk_2,\vk_3)$ to the 
three-point cross-correlation of interest.
We find that the leading terms (as $\ee \to 0$) in the different 
contributions are given by
%%%%%%%%%%%%%%%%%%%%%%%%%%%%%%%%%%%%%%%%%%%%%%%%%%%%%%%%%%%%%%%%%%%%%%%%%%%%%%%
\begin{subequations}
\begin{align}
\cB_1(\vk_1,\vk_2,\vk_3) 
&= -\f{3\,\HI^6\, \ei^2}{{32\, \Mpl^2\,k_1^3\, k_2^3\, k_3^3\,\eai}}
\l(\f{\ei}{{\ee}}\r)^4\, (k_1^2-k_2^2-k_3^2)\, 
\l[3\, k_1^3+4\,(k_2^3+k_3^3)\r],\\
\cB_2(\vk_1,\vk_2,\vk_3)
&= \f{3\,\HI^6\,\ei^2}{32\, \Mpl^2\,k_1^3\, k_2^3\, k_3^3\,\eai}\,
\l(\f{\ei}{\ee}\r)^4\,(k_1^2-k_2^2-k_3^2)\,
\l[k_1^3-4\, (k_2^3+k_3^3)\r],\\
\cB_3(\bm{k}_1,\bm{k}_2,\bm{k}_3) 
&= \f{9\,\HI^6\,\ei^2}{128\, \Mpl^2\,\eai\, k_1^3\, k_2^3\, k_3^5}\,
\l(\f{\ei}{\ee}\r)^4\,
\l[(k_1^2-k_2^2)^2+2\, (k_1^2+k_2^2)\,k_3^2-3\, k_3^4\r]\,
(3\,k_1^3+4\,k_2^3) + \vk_2 \leftrightarrow\vk_3,\\
\cB_4(\vk_1,\vk_2,\vk_3)
&=\f{81\,\pi\,\HI^6}{64\,\Mpl^2\,k_1^5\,k_2^3\,k_3^5\,\ee^3}\,
\l[(k_1-k_2)^2+2\,(k_1^2+k_2^2)\,k_3^2-3\,k_3^4\r]+ \vk_2 \leftrightarrow\vk_3,\\
\cB_5(\vk_1,\vk_2,\vk_3)
&=  \f{9\,\HI^6}{64\, \Mpl^2\, k_2^5\, k_3^5\,\eai}\, 
\l(\f{\ei}{\ee}\r)^6\,
\l[k_1^4+k_2^4-2\, k_1^2\,(k_2^2+k_3^2)+6\, k_2^2\, k_3^2+k_3^4\r],\\
\cB_6(\vk_1,\vk_2,\vk_3)
&=  \f{63\,\HI^6}{64\,\Mpl^2\, k_2^5\, k_3^5\,\eai}\,
\l(\f{\ei}{\ee}\r)^6\,
\l[k_1^4+k_2^4-2\, k_1^2\, (k_2^2+k_3^2)+6\, k_2^2\, k_3^2+k_3^4\r].
\end{align}
\end{subequations}
%%%%%%%%%%%%%%%%%%%%%%%%%%%%%%%%%%%%%%%%%%%%%%%%%%%%%%%%%%%%%%%%%%%%%%%%%%%%%%%
Clearly, among the different contributions, it is the fifth and the 
sixth terms, viz.~${\cal B}_5(\vk_1,\vk_2,\vk_3)$ 
and~${\cal B}_6(\vk_1,\vk_2,\vk_3)$, that dominate.
Note that these contributions behave as~$(\ei/\ee)^6$, just as the
scalar power spectrum~$\ps(k)$ does when $p=6$ [cf. Eq.~\eqref{eq:ps-p6}].
Since the magnetic power spectrum~$\pb(k)$ is a constant [cf. 
Eq.~\eqref{eq:pbe-nm4}], we find that resulting non-Gaussianity
parameter~$\bnl(\vk_1,\vk_2,\vk_3)$ turns out to be independent
of~$\ee$ and is given by
\begin{equation}
\bnl(\vk_1,\vk_2,\vk_3)
= \f{k_1^3}{2\, k_2^2\, k_3^2\, (k_2^3+k_3^3)}\,
\l[k_1^4+k_2^4 -2\, k_1^2\, (k_2^2+k_3^2)+6\, k_2^2\,k_3^2+k_3^4\r].
\label{eq:bnl-p6}
\end{equation}
\end{widetext}
%%%%%%%%%%%%%%%%%%%%%%%%%%%%%%%%%%%%%%%%%%%%%%%%%%%%%%%%%%%%%%%%%%%%%%%%%%%%%%%

Let us now discuss the amplitude of the non-Gaussianity
parameter~$\bnl(\vk_1,\vk_2,\vk_3)$ for various configurations of 
the wave vectors.
In the equilateral limit wherein $k_1=k_2=k_3=k$, we find that the
parameter $\bnl(k)$ given by Eq.~\eqref{eq:bnl-p6} above turns out
to be independent of the wave number and is given by
\begin{equation}
\bnl^{\mathrm{eq}}(k) \simeq \f{5}{4}.
\label{eq:bnl-p6-eq}
\end{equation}
In the flattened limit with the configuration $k_1=2\,k_2=2\,k_3=2\, 
k$, we find that 
\begin{equation}
\bnl^\mathrm{fl}(k) \simeq 16.
\label{eq:bnl-p6-fl}
\end{equation}
Finally, in the squeezed limit wherein $k_1 \to 0$ and $k_2=k_3
=k$, we obtain that
\begin{align}
\bnl^\mathrm{sq}(k_1, k) 
&\simeq 2 \l(\f{k_1}{k}\r)^3 
\Biggl\{1 - \f{1}{2}\,\l(\f{k_1}{k}\r)^2\nn\\ 
&\quad+ \mathcal{O}\l[\l(\f{k_1}{k}\r)^4\r]\Biggr\}
+ \mathcal{O}(k^2\, \ee^2).\label{eq:bnl-p6-sq}
\end{align}
Evidently, in the squeezed limit, the leading contribution to 
$\bnl(k_1,k)$ is proportional to~$k_1^3$, which vanishes. 
However, we find that there is a subleading contribution to the 
non-Gaussianity parameter which is proportional to $(k^2\, \ee^2)$.
But, since $(k\,\ee)$ is very small, this contribution is also rather
small. 
For instance, if we choose $k=k_\ast \simeq 5\times 10^{-2}\,\mpcinv$, 
then, for the choice of $\ee=-10^{-19}\,\mathrm{Mpc}$, we obtain that 
$\bnl^\mathrm{sq} \simeq \mathcal{O}(10^{-42})$.
We shall comment further about the behavior of the non-Gaussianity 
parameter in the squeezed limit when we discuss the consistency 
relation in the subsequent subsection.

\begin{figure*}
\centering
\includegraphics[scale=0.3]{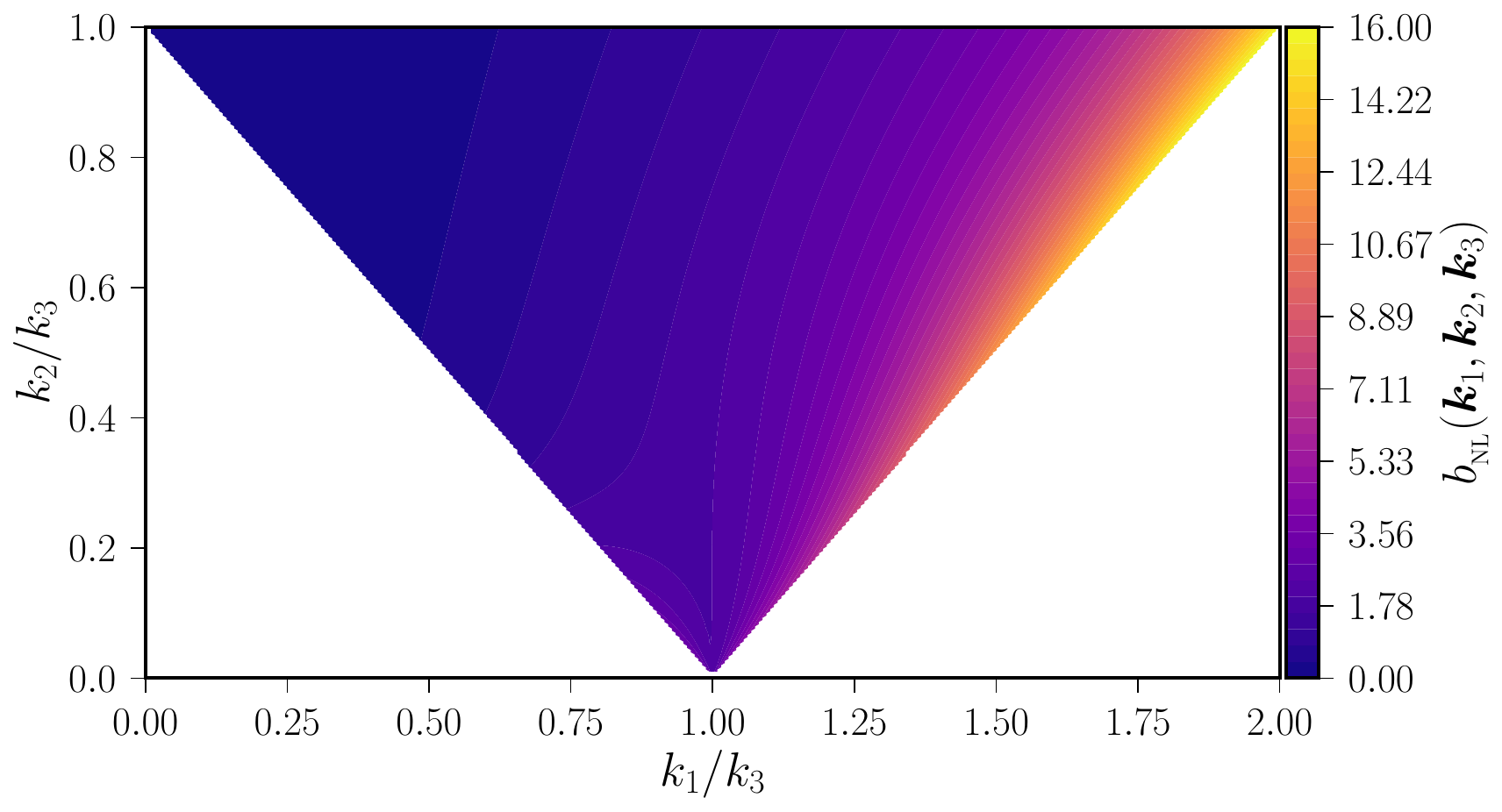}
\includegraphics[scale=0.3]{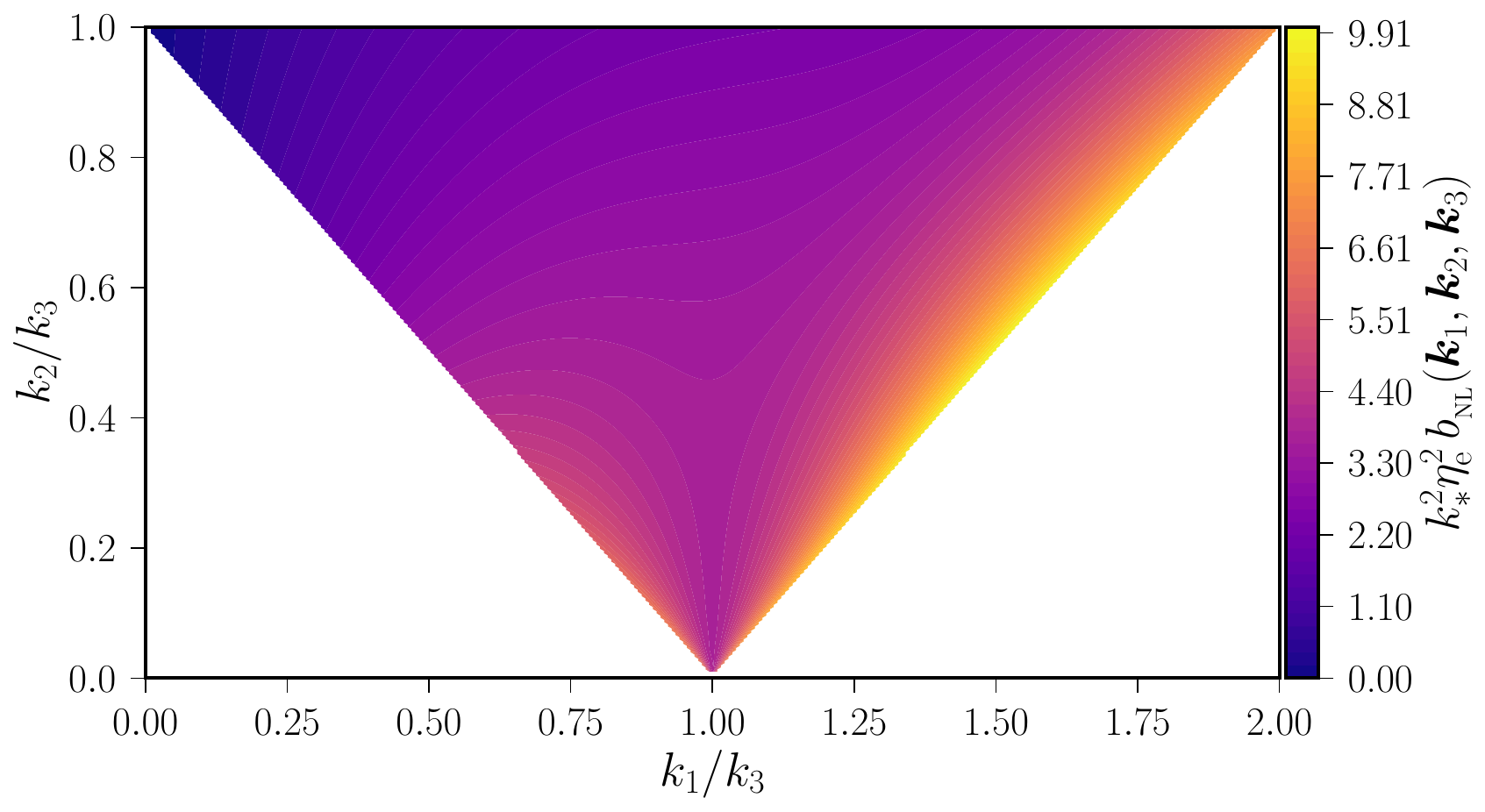}
\caption{We present the complete structure of the non-Gaussianity
parameter~$\bnl(\vk_1,\vk_2,\vk_3)$ as density plots in the plane 
of $k_1/k_3$-$k_2/k_3$ for the cases of $p=6$ (on the left) and 
$p=4$ (on the right). 
Note that, in the case of $p=4$, due to the behavior $\bnl(\vk_1,\vk_2,\vk_3) 
\propto \ee^{-2}$, we have plotted the dimensionless 
combination~$k_3^2\,\ee^2\,\bnl(\vk_1,\vk_2,\vk_3)$.
We have set $\eai = 10^{-3}$, $\ei = -10^6\,\mathrm{Mpc}$ and $\ee = 
-10^{-19}\, \mathrm{Mpc}$, as discussed in the text.
Further, we have set $k_3=k_\ast=5\times 10^{-2}\,\mpcinv$ to arrive
at the plots.
We should highlight that, in both the cases of $p=6$ and $p=4$, the
magnitude of $\bnl(\vk_1,\vk_2,\vk_3)$ spans over a wide range of 
values between the flattened limit (on the top right corners of the 
density plots wherein $k_1=2\,k_2 = 2\,k_3$) and the squeezed limit
(on the top left corners wherein $k_1 \to 0$ and $k_2 \simeq k_3$).}
\label{fig:bnl-2D}
\end{figure*}

To understand the complete structure of the non-Gaussianity parameter
$\bnl(\vk_1,\vk_2,\vk_3)$, in Fig.~\ref{fig:bnl-2D}, we have illustrated
it for all possible configurations of wave vectors.
%%%%%%%%%%%%%%%%%%%%%%%%%%%%%%%%%%%%%%%%%%%%%%%%%%%%%%%%%%%%%%%%%%%%%%%%%%%%%%%
%%%%%%%%%%%%%%%%%%%%%%%%%%%%%%%%%%%%%%%%%%%%%%%%%%%%%%%%%%%%%%%%%%%%%%%%%%%%%%%
We have presented the behavior of $\bnl(\vk_1,\vk_2,\vk_3)$ for
$p=6$ as a density plot in the two-dimensional plane 
of $k_1/k_3$-$k_2/k_3$, with $k_3=k_\ast$. 
Note that this plot is obtained using the complete expression for
$\bnl(\vk_1,\vk_2,\vk_3)$ as given in Eq.~\eqref{eq:bnl-p6}.
We find that $\bnl(\vk_1,\vk_2,\vk_3)$ peaks in the region around
the flattened limit (i.e. when $k_1 = k_3 + k_2$) where it attains 
the maximum value of~$16$.
It steadily falls towards zero in the squeezed limit, i.e. when $k_1 
\to 0$ and $k_2 = k_3$.
These behavior are indeed expected from the results in the flattened
and squeezed limits we have already arrived at in 
Eqs.~\eqref{eq:bnl-p6-fl} and~\eqref{eq:bnl-p6-sq}.

%%%%%%%%%%%%%%%%%%%%%%%%%%%%%%%%%%%%%%%%%%%%%%%%%%%%%%%%%%%%%%%%%%%%%%%%%%%%%%%

\begin{widetext}

\subsubsection{Case of $p=4$}

When $p=4$, as in the previous case, we can utilize the behavior 
of~$\epsilon_1(\eta)$ and~$J(\eta)$ as well as the associated solutions
for the Fourier mode functions~$f_k(\eta)$ and $\bar{A}_k(\eta)$ [as 
given in Eqs.~\eqref{eq:fk-p4}] and~\eqref{eq:Abk}] to evaluate the 
integrals~$\cG_C(\vk_1,\vk_2,\vk_3)$ and arrive at the different 
contributions~$\cB_C(\vk_1,\vk_2,\vk_3)$ to the cross-correlation.
Again, we choose the value of~$n$ such that $n\,p=-4$, which leads 
to $n=-1$ in this case.
We obtain the different contributions~$\cB_C(\vk_1,\vk_2,\vk_3)$ to be
%%%%%%%%%%%%%%%%%%%%%%%%%%%%%%%%%%%%%%%%%%%%%%%%%%%%%%%%%%%%%%%%%%%%%%%%%
\begin{subequations}    
\begin{align}
\cB_1(\vk_1,\vk_2,\vk_3) 
&=-\f{9\,\HI^6\,\ei^2}{16\,\Mpl^2\, k_2^3\, k_3^3\,\eai}\,
\l(\f{\ei}{\ee}\r)^2\, \l(k_1^2-k_2^2-k_3^2\r),\\
\cB_2(\bm{k}_1,\bm{k}_2,\bm{k}_3) 
&= \f{9\,\HI^6\,\ei^2}{16\,\Mpl^2\, k_2^3\, k_3^3\,\eai}\,
\l(\f{\ei}{\ee}\r)^2\, \l(k_1^2-k_2^2-k_3^2\r),\\
\cB_3(\vk_1,\vk_2,\vk_3)
&=\f{27\, \HI^6\,\ei^2}{64\,\Mpl^2\,k_2^3\, k_3^5\,\eai}\,
\l(\f{\ei}{\ee}\r)^2\,
\l[(k_1^2-k_2^2)^2+2\,(k_1^2+k_2^2)\,k_3^2-3\, k_3^4\r]
+\vk_2 \leftrightarrow \vk_3,\\
\cB_4(\vk_1,\vk_2,\vk_3)
&=\f{27\,\pi\,\HI^6}{64\,\Mpl^2\,k_1^3\,k_2^3\,k_3^5\,\ee}
\l[(k_1^2-k_2^2)^2+2\,(k_1^2+k_2^2)\,k_3^2-3\, k_3^4\r]
+\vk_2 \leftrightarrow \vk_3,\\
\cB_5(\vk_1,\vk_2,\vk_3)
&= \f{27\, \HI^6}{128\,\Mpl^2\, k_2^5\, k_3^5\,\eai}\,
\l(\f{\ei}{\ee}\r)^4\,
\l[k_1^4+k_2^4-2\, k_1^2\, (k_2^2+k_3^2)
+6\, k_2^2\, k_3^2+k_3^4\r],\\
\cB_6(\vk_1,\vk_2,\vk_3)
&=\f{243\, \HI^6}{128\, \Mpl^2\,k_2^5\, k_3^5\,\eai}\,
\l(\f{\ei}{\ee}\r)^4\, \l[k_1^4+k_2^4-2\, k_1^2\, (k_2^2+k_3^2)
+6\, k_2^2\, k_3^2+k_3^4\r].
\end{align}
\end{subequations}
%%%%%%%%%%%%%%%%%%%%%%%%%%%%%%%%%%%%%%%%%%%%%%%%%%%%%%%%%%%%%%%%%%%%%%%%% 
Note that, as in the case of~$p=6$, it is the fifth and the sixth 
terms, viz.~$\cB_5(\vk_1,\vk_2,\vk_3)$ and~$\cB_6(\vk_1,\vk_2,\vk_3)$,
that dominate the contributions.
If we take into account these contributions, we find that the corresponding 
non-Gaussianity parameter~$\bnl(\vk_1,\vk_2,\vk_3)$ turns out to be
\begin{align}
\bnl(\vk_1,\vk_2,\vk_3) 
&= \f{15\, k_1}{16\, k_2^2\, k_3^2\, (k_2^3+k_3^3)\,\ee^2}\,
\l[k_1^4+k_2^4-2\, k_1^2\, (k_2^2+k_3^2)+6\, k_2^2\, k_3^2+k_3^4\r]. 
\label{eq:bnl-full-p4} 
\end{align}
%%%%%%%%%%%%%%%%%%%%%%%%%%%%%%%%%%%%%%%%%%%%%%%%%%%%%%%%%%%%%%%%%%%%%%%%%
\end{widetext}

A notable feature of the non-Gaussianity parameter~$\bnl(\vk_1,\vk_2,\vk_3)$ 
in this case is the dependence on~$\ee$ as~$\ee^{-2}$. 
This is a large factor and it affects the amplitude of the parameter
across all configurations.
On inspecting the various limits, we find that, in the equilateral limit
(i.e. when $k_1=k_2=k_3=k$), the expression Eq.~\eqref{eq:bnl-full-p4} above
simplifies to 
\begin{equation}
\bnl^\mathrm{eq}(k) \simeq \f{75}{32\, k^2\, \ee^2}.
\end{equation}
For the value of the parameter $\ee=-10^{-19}\,\mathrm{Mpc}$ as fixed earlier, 
at $k_\ast=5\times 10^{-2}\,\mathrm{Mpc}^{-1}$, we obtain that~$\bnl^\mathrm{eq} 
\simeq 9.375\times 10^{40}$.
In the flattened limit (i.e. when $k_1=2\,k_2=2\,k_3=2\, k$), we get
\begin{equation}
\bnl^\mathrm{fl}(k) \simeq \f{15}{2\,k^2\,\ee^2},
\label{eq:bnl-p4-fl}
\end{equation}
which at $k_\ast$ turns out to be $\bnl^\mathrm{fl}\simeq 3\times 10^{41}$.
In the squeezed limit (i.e. when $k_1 \to 0$ and $k_2=k_3=k$), we obtain
that
\begin{align}
\bnl^\mathrm{sq}(k_1, k) 
&\simeq \f{15}{4\,k^2\, \ee^2}\, \l(\f{k_1}{k}\r) 
\biggl\{1 - \f{1}{2}\l(\f{k_1}{k}\r)^2\nn\\
&\quad+ {\cal O}\l[\l(\f{k_1}{k}\r)^4\r]\biggr\}
+ \mathcal{O}(k\, \ee).\label{eq:bnl-p4-sq}
\end{align}
We can see that, as $k_1 \to 0$ in the squeezed limit, the leading order 
contribution to $\bnl(k_1,k)$ is proportional to $k_1$ and hence vanishes. 
However, there is a subleading contribution that is proportional to $(k\, 
\ee)$, which becomes relevant only in the squeezed limit. 
For the pivot scale of $k_\ast=5\times 10^{-2}\,\mpcinv$ and, for our 
choice of~$\ee=-10^{-19}\,\mathrm{Mpc}$, the contribution proves to be 
very small. 
In fact, we find that $\bnl^\mathrm{sq} \simeq \mathcal{O}(10^{-21})$.
We shall discuss the corresponding implications for the consistency
relation governing the three-point function in the next subsection.

As in the earlier case, when $p=4$, we have presented the complete structure 
of the  parameter~$\bnl(\vk_1,\vk_2,\vk_3)$ as a density plot in Fig.~\ref{fig:bnl-2D}.
Note that this plot is obtained using the expression for 
$\bnl(\vk_1,\vk_2,\vk_3)$ as given in Eq.~\eqref{eq:bnl-full-p4}.
We find that $\bnl(\vk_1,\vk_2,\vk_3)$ peaks in the region around the 
flattened limit where $k_1 = k_2 + k_3$, as in the $p=6$ case.
In the case of $p=4$, the parameter reaches extremely large values across 
all configurations because of the general behavior of $\bnl(\vk_1,\vk_2,
\vk_3)$ as $\ee^{-2}$, as we mentioned earlier.
Hence, we have plotted the combination~$k_3^2\,\ee^2\,\bnl(\vk_1,\vk_2,\vk_3)$ 
to better understand its structure for different configurations of wave numbers.
This combination reaches values as large as $7.5$ in the flattened limit and 
it vanishes in the squeezed limit.
These behavior are as expected from the different limits we already arrived 
at in Eqs.~\eqref{eq:bnl-p4-fl} and~\eqref{eq:bnl-p4-sq}.
Such large values for the non-Gaussianity parameter~$\bnl(\vk_1,\vk_2,\vk_3)$ 
may lead to strong non-Gaussian corrections to the power spectrum of the 
magnetic field~$\pb(k)$. 
We shall relegate the examination of such a possibility to App.~\ref{app:pc-bnl},
wherein we shall discuss some interesting results.

%%%%%%%%%%%%%%%%%%%%%%%%%%%%%%%%%%%%%%%%%%%%%%%%%%%%%%%%%%%%%%%%%%%%%%%%%%%%%%%

\subsection{Consistency relation}

The behavior of the three-point cross-correlation $\cB(\vk_1,\vk_2,\vk_3)$
in the squeezed limit (i.e. when $k_1 \to 0$ and $\vk_2 = -\vk_3$) is 
expected to be governed by the consistency relation that we had mentioned
earlier.
This relation allows us to express the non-Gaussianity 
parameter~$\bnl(\vk_1,\vk_2,\vk_3)$ in the squeezed 
limit in terms of the spectral index of the magnetic 
power spectrum, say, $\nb(k)$, which is, in general, 
defined as $\nb(k)=\d \ln\pb(k)/\d \ln k$.
In the case of slow roll inflation, the consistency relation is obtained 
to be~\cite{Jain:2012ga,Nandi:2021lpf}
\begin{eqnarray}
\bnl^{\mathrm{cr}} = \f{4-\nb}{2}.\label{eq:bnl-cr}
\end{eqnarray}
This relation has been verified by explicit calculations in the case of slow 
roll inflation [in this context, see Eqs.~\eqref{eq:bnl-dS-n2} 
and~\eqref{eq:bnl-dS-n1} as well as the 
associated discussion in App.~\ref{app:DC}]. 
Both the ultra slow roll scenarios described so far have the coupling function
behaving as $J \propto \eta^{-2}$, thereby leading to a scale invariant power 
spectrum for the magnetic field (i.e. $\nb=0$). 
As should be evident from Eqs.~\eqref{eq:bnl-p6-sq}, \eqref{eq:bnl-p4-sq}
and~\eqref{eq:bnl-cr}, the consistency relation is violated in these cases. 
Besides the ultra slow roll scenarios discussed thus far which lead to a 
scale invariant~$\pb(k)$, in App.~\ref{app:DC}, we have considered another 
scenario where $\pb(k)$ is not scale invariant, but has a spectral index 
of $\nb=2$. 
In this case, according to the consistency relation, we should obtain that 
$\bnl^\mathrm{cr}=1$. 
However, we instead obtain~$\bnl^\mathrm{sq} = 0$
[cf. Eq.~\eqref{eq:bnl-app-sq}], which also violates the consistency 
relation.

In Tab.~\ref{tab:one}, we have listed the values of $\bnl^\mathrm{sq}$ 
obtained in the different scenarios we have considered. 
%%%%%%%%%%%%%%%%%%%%%%%%%%%%%%%%%%%%%%%%%%%%%%%%%%%%%%%%%%%%%%%%%%%%%%%%%%%%%%%
\begin{table}[t]
\begin{tabular}{|c|c|c|c|c|}
\hline
$\alpha$ & $\nb$ & $\bnl^\mathrm{sq}$ in
& $\bnl^\mathrm{sq}$ in & $\bnl^\mathrm{cr}$\\
& & slow roll & ultra slow roll & \\
\hline
$2$ & $0$ & $2$ & $0$ & $2$\\
\hline
$1$ & $2$ & $1$ & $0$ & $1$\\ 
\hline
\end{tabular}
\caption{We have listed the values of the non-Gaussianity 
parameter~$\bnl(\vk_1,\vk_2,\vk_3)$ 
obtained in the squeezed limit in the various scenarios of interest.
We have assumed that $J \propto \eta^{-\alpha}$ in arriving at these results.
We have also listed the corresponding values of $\bnl$ that we expect to
obtain from the consistency condition given in Eq.~\eqref{eq:bnl-cr}.
It is clear that, while the consistency relation is satisfied in slow roll
inflation, it is violated in the ultra slow roll scenarios.}\label{tab:one}
\end{table}
%%%%%%%%%%%%%%%%%%%%%%%%%%%%%%%%%%%%%%%%%%%%%%%%%%%%%%%%%%%%%%%%%%%%%%%%%%%%%%%
We have also listed the values of $\bnl^\mathrm{cr}$ that we are expected
to obtain.
From the table, we can clearly see that, while the consistency 
relation is satisfied in slow roll inflation, for a similar behavior 
of~$J(\eta)$, it is violated in ultra slow roll inflation.
The standard consistency relations arise in situations wherein the amplitude 
of the long wave length mode freezes~\cite{Maldacena:2002vr,Seery:2005wm,
Cheung:2007sv,Senatore:2012wy}.
In slow roll inflation, we find that the consistency relation in Eq.~\eqref{eq:bnl-cr}
is satisfied by the non-Gaussianity parameter~$\bnl(\vk_1,\vk_2,\vk_3)$ in the
squeezed limit since, in such situations, the amplitude of the curvature 
perturbation goes to a constant value on super-Hubble scales.
In contrast, in pure ultra slow roll inflation, the amplitude of the curvature
perturbations grows indefinitely at late times. 
The violation of the consistency relation [as given by Eq.~\eqref{eq:bnl-cr}] 
that we encounter in the ultra slow roll scenarios can be ascribed to the growth 
in the amplitude of the curvature perturbations on super-Hubble scales.
One expects the consistency relation in Eq.~\eqref{eq:bnl-cr} to be modified in 
such situations (for related discussions in the case of the scalar bispectrum, 
see Refs.~\cite{Bravo:2020hde,Suyama:2020akr,Suyama:2021adn}).
It will be interesting to derive a generic consistency condition that governs 
the non-Gaussianity parameter~$\bnl(\vk_1,\vk_2,\vk_3)$ in the squeezed limit,
including non-trivial evolution of the long wave length curvature perturbation.
However, such a derivation is beyond the scope of this work and we hope to 
address the problem in the future.

%%%%%%%%%%%%%%%%%%%%%%%%%%%%%%%%%%%%%%%%%%%%%%%%%%%%%%%%%%%%%%%%%%%%%%%%%%%%%%%

\section{Conclusions}\label{sec:conc}

Models of inflation that permit a brief phase of ultra slow roll have 
attracted attention in the recent literature because they lead to scalar 
power spectra with higher amplitudes over small scales. 
Such scalar spectra with enhanced power on small scales can produce copious
amounts of primordial black holes and also induce secondary gravitational
waves of considerable strengths.
In this work, we investigated the generation of magnetic fields in scenarios
of ultra slow roll inflation.
It has been pointed out that there arise certain challenges in generating
magnetic fields with nearly scale invariant spectra and of observed strengths
in single field inflationary models that permit a short period of ultra slow
roll (see Ref.~\cite{Tripathy:2021sfb}; in this context, also see
Ref.~\cite{Tripathy:2022iev}).
To circumvent such a challenge, rather than considering the electromagnetic
action involving the popular non-conformal coupling function $J(\phi)$ 
[cf. Eq.~\eqref{eq:ema-1}], we worked with an action involving a function
of the form~$J(X)$ [cf. Eq.~\eqref{eq:ema-2}].
We focused on {\it pure}\/ ultra slow roll scenarios.
In other words, we assumed that the phase of ultra slow roll lasts throughout 
the entire duration of inflation.
While such scenarios do not seem viable when compared with the observations, 
we considered these scenarios to gain a better understanding of the effects 
of ultra slow roll on the two-point and three-point functions involving the 
magnetic fields.
We worked with a non-conformal coupling function~$J(X)$ which turned out to 
be proportional to a power of the first slow roll parameter in the ultra slow
roll scenarios.
We find that, for suitable choices of parameters describing the non-conformal
coupling function and the inflationary scenario, we can obtain scale invariant
spectra for the magnetic fields.
However, we should mention that, for our choice of parameters, the coupling 
function does indeed exhibit the strong coupling problem during early times.
One of the ways to ameliorate this issue is to consider a `sawtooth' coupling 
(for detailed discussions, see, for example, Refs.~\cite{Ferreira:2013sqa,
Nandi:2021lpf,Cecchini:2023bqu}).
We are currently studying implementations of such coupling functions in ultra 
slow roll scenarios.
\color{black}

For our choice of~$J(X)$, we also calculated the three-point cross-correlation 
$\cB(\vk_1,\vk_2,\vk_3)$ involving the curvature perturbation and the magnetic
fields in the ultra slow roll scenarios.
We examined the amplitude and shape of the associated non-Gaussianity parameter
$\bnl(\vk_1,\vk_2,\vk_3)$ arising in a few different cases.
When compared to slow roll inflation, we find that the values of the 
non-Gaussianity parameter in the ultra slow roll scenario of $p=6$, 
with $n=-2/3$, leading to scale invariant scalar and magnetic power 
spectra, are considerably smaller [cf. Eqs.~\eqref{eq:bnl-p6-eq}, 
\eqref{eq:bnl-p6-fl} and \eqref{eq:bnl-p6-sq}].
However, we find that the values of $\bnl(\vk_1,\vk_2,\vk_3)$ can vary 
significantly depending on the ultra slow roll scenario and the value 
of~$n$ being considered.
For example, in the case wherein we obtain a scale invariant spectrum for
the magnetic field and a blue scalar power spectrum [with $\ps(k) \propto
k^2$], we obtain very large values for the non-Gaussianity parameter (see 
right panel of Fig.~\ref{fig:bnl-2D} and the accompanying discussion).
In contrast, in the ultra slow roll scenario which leads to a scale 
invariant scalar power spectrum and a blue magnetic power spectrum 
[with $\pb(k)\propto k^2$], we obtain very small values for the 
non-Gaussianity parameter (see App.~\ref{app:DC}).
Specifically, we had examined the validity of the consistency condition 
that the three-point function of interest is expected to satisfy in the 
squeezed limit [cf. Eq.~\eqref{eq:bnl-cr}].
This consistency relation is known to be valid in the standard slow 
roll inflationary scenarios (see App.~\ref{app:DC} for a brief discussion
on this point).
But, we observe that, in all the ultra slow roll scenarios that we have
considered, the aforementioned consistency relation is violated.
This can be attributed to the fact that, in the ultra slow roll scenarios, in 
contrast to slow roll inflation, the amplitude of the curvature perturbation 
grows strongly on super-Hubble scales.

It would be interesting to investigate the manner in which the super-Hubble
evolution of the long wavelength mode (the curvature perturbation in this 
case) modifies the consistency relation (for discussions on the scalar 
bispectrum in such situations, see, for instance, Refs.~\cite{Bravo:2020hde,
Suyama:2021adn}).
Rather than the pure ultra slow roll scenarios, it would be compelling to study
the production of magnetic fields in inflationary models involving a brief 
phase of ultra slow roll sandwiched by two epochs of slow roll inflation.
Such scenarios are in better agreement with the observations and would hence 
provide us with testable predictions.
We defer these interesting possibilities to a future work.

%%%%%%%%%%%%%%%%%%%%%%%%%%%%%%%%%%%%%%%%%%%%%%%%%%%%%%%%%%%%%%%%%%%%%%%%%%%%%%

\acknowledgements

ST thanks the Indian Institute of Astrophysics, Bengaluru, India, for 
financial support through a postdoctoral fellowship.
DC would like to thank the Indian Institute of Astrophysics, Bengaluru,
India, and the Department of Science and Technology, Government of India, 
for support through the INSPIRE Faculty Fellowship
grant~DST/INSPIRE/04/2023/000110.
DC would also like to thank the Indian Institute of Science, Bengaluru, 
India, for support through the C.~V.~Raman postdoctoral fellowship. 
HVR thanks the Raman Research Institute, Bengaluru, India, for support 
through a postdoctoral fellowship.
HVR also acknowledges support by the MUR PRIN2022 Project “BROWSEPOL:~Beyond 
standaRd mOdel With coSmic microwavE background POLarization”-2022EJNZ53 
financed by the European Union---Next Generation EU.
LS wishes to thank the Indo-French Centre for the Promotion of Advanced
Research for support of the proposal 6704-4 under the Collaborative 
Scientific Research Programme.

%%%%%%%%%%%%%%%%%%%%%%%%%%%%%%%%%%%%%%%%%%%%%%%%%%%%%%%%%%%%%%%%%%%%%%%%%%%%%%

\appendix

\section{Does the energy density of the electromagnetic field remain
positive?}\label{app:rho}

Earlier, in the case of the action in Eq.~\eqref{eq:ema-2}, which contains
the non-conformal coupling of the form $J^2(X) \simeq J_0\, X^{n/2}$, we had 
obtained the corresponding energy density~$\rem$ of the electromagnetic field 
to be given by Eq.~\eqref{eq:rem-2}.
While the expectation values~$\langle \hat{\rho}_{_{\mathrm{B}}}\rangle$
and~$\langle \hat{\rho}_{_{\mathrm{E}}}\rangle$ are clearly positive 
definite [cf. Eqs.~\eqref{eq:rbe}, \eqref{eq:pbe-d} and~\eqref{eq:pbe}], 
the quantity $\langle \hat{\rho}_{_\mathrm{{EM}}} \rangle$ contains~$n$ 
which may have either sign.
This may raise the concern as to whether the total energy
density of the electromagnetic field remains positive definite.
To address the concern, let us study the behavior of the energy density 
of the electromagnetic field over the entire span of conformal time of 
our interest.
For our choice of the non-conformal coupling function, i.e. 
$J=(\eta/\ee)^{n\,p/2}$ [cf. Eqs.~\eqref{eq:Jf} and~\eqref{eq:e1-p}],
we obtain the quantity~$\cA_k$ describing the Fourier modes of the 
electromagnetic vector potential to be
\begin{equation}
\cA_k(\eta) = \sqrt{-\f{\pi\, \eta}{4}}\,
\mathrm{e}^{i\,[-(n\,p/2)+1]\pi/2}\, H^{(1)}_{-(n\,p-1)/2}(-k\,\eta),
\end{equation}
where $H_\nu^{(1)}(z)$ denotes the Hankel functions of the first kind.
On substituting this mode function in Eq.~\eqref{eq:pbe}, we obtain 
the power spectra of the magnetic and electric fields to be
\begin{align}
\pb(k,\eta) &= \f{k^5}{2\,\pi^2\,a^4}\,
\l(-\f{\pi\, \eta}{4}\r)\,
\l\vert H^{(1)}_{-(n\,p-1)/2}(-k\,\eta)\r\vert^2,\quad\\
\pe(k,\eta) &= \f{k^5}{2\,\pi^2\,a^4}\,
\l(-\f{\pi\, \eta}{4}\r)\,
\l\vert H^{(1)}_{-(n\,p+1)/2}(-k\,\eta)\r\vert^2.\quad
\end{align}
The expectation value of the total energy density of the electromagnetic 
field can be expressed as [cf. Eq.~\eqref{eq:rem-2}]
\begin{equation}
\langle \hat{\rho}_{_{\mathrm{EM}}}(\eta)\rangle
= \int_{k_{\mathrm{min}}}^{k_{\mathrm{max}}} 
\d \,{\rm ln}\,k\,
\langle \hat{\rho}_{_{\mathrm{EM}}}(k,\eta)\rangle,
\end{equation}
where the expectation value of the energy density~$\rem(k,\eta)$ 
associated with a particular mode, say $k$, is given by
\begin{equation}
\langle \hat{\rho}_{_{\mathrm{EM}}}(k,\eta)\rangle
= (1+2\,n)\,\pe(k,\eta)+ (1-2\,n)\,\pb(k,\eta),
\label{eq:rhoem-total}
\end{equation}
while $k_{\mathrm{min}}$ and $k_{\mathrm{max}}$ denote the smallest
and largest wave numbers of observational interest.
In our case, the expectation value of the energy density of the 
electromagnetic field associated with a particular mode, say $k$, 
is given by
\begin{align}
\langle \hat{\rho}_{_{\mathrm{EM}}}(k,\eta)\rangle
&= \f{k^5\,\eta}{8\,\pi^2\,a^4}\,
\biggl\{\l\vert H^{(1)}_{-(n\,p+1)/2}(-k\,\eta)\r|^2 \nn\\
&\quad+ \l\vert H^{(1)}_{-(n\,p-1)/2}(-k\,\eta)\r\vert^2\nn\\ 
&\quad+ 2\, n\,\biggl[\l\vert H^{(1)}_{-(n\,p+1)/2}(-k\,\eta)\r\vert^2\nn\\ 
&\quad- \l\vert H^{(1)}_{-(n\,p-1)/2}(-k\,\eta)\r\vert^2\biggr]\biggr\}.
\label{eq:rhoem-hankel}
\end{align}

Let us now evaluate the above spectral energy density in the three 
ultra slow roll scenarios of interest. 
For the case when $p=6$ and $n=-2/3$, we obtain that
\begin{equation}
\langle \hat{\rho}_{_{\mathrm{EM}}}(k,\eta)\rangle
= \f{\HI^4}{12\, \pi^2}\,
(6\, k^4\, \eta^4 +20\, k^2\, \eta^2 +63).
\end{equation}
For the second case, i.e. when $p=4$ and $n=-1$, we obtain that
\begin{equation}
\langle \hat{\rho}_{_{\mathrm{EM}}}(k,\eta)\rangle
= \f{\HI^4}{4\, \pi^2}\,
(2\, k^4\, \eta^4 +8\, k^2\, \eta^2 +27).
\end{equation}
Finally, for the case wherein $p=6,\,n=-1/3$ (discussed in 
App.~\ref{app:DC}), we find that
\begin{equation}
\langle \hat{\rho}_{_{\mathrm{EM}}}(k,\eta)\rangle
= \f{\HI^4}{12\, \pi^2}\,  k^2 \eta^2\,
(6\, k^2\, \eta^2 +5).
\end{equation}
From the above three expressions, it is clear that the spectral energy 
density of the electromagnetic 
field~$\langle \hat{\rho}_{_{\mathrm{EM}}}(k,\eta)\rangle$ is positive 
at all times for all the three cases. 
Therefore, the total electromagnetic energy density remains positive 
over the full span of time in all the three ultra slow roll scenarios 
of our interest.

%%%%%%%%%%%%%%%%%%%%%%%%%%%%%%%%%%%%%%%%%%%%%%%%%%%%%%%%%%%%%%%%%%%%%%%%%%%%%%%

\begin{widetext}
\section{The integrals}\label{app:bnl-calG}

In Sec.~\ref{sec:calB}, we presented the results for the different 
contributions~$\cB_C(\vk_1,\vk_2,\vk_3)$ to the three-point cross-correlation 
of interest in the scenario of pure ultra slow roll inflation for the cases 
of $p=6$ and $p=4$.
We had considered the situation wherein $n\,p=-4$, which leads to a scale 
invariant spectrum for the magnetic field.
To arrive at the different contributions, we needed to evaluate the integrals 
$\cG_C(\vk_1,\vk_2,\vk_3)$ [defined in Eqs.~\eqref{eq:cG}] which depended on 
the background quantities such as the scale factor~$a$, the Hubble parameter~$H$,
the first slow roll parameter~$\epsilon_1$, and the mode functions~$f_k$ 
and~$\bar{A}_k$ that describe the curvature perturbations and the 
electromagnetic vector potential.
Note that, we assume the scale factor to be of the de Sitter form with a 
constant Hubble parameter~$\HI$, while the first slow parameter $\epsilon_1$ 
is given by Eq.~\eqref{eq:e1-p}.
Moreover, the scalar mode functions~$f_k$ are given by Eqs.~\eqref{eq:fk-p6} 
and~\eqref{eq:fk-p4} (in the cases of $p=6$ and $p=4$, respectively), whereas
the mode function associated with the electromagnetic vector potential $\bar{A}_k$
is given by Eq.~\eqref{eq:Abk}.
In this Appendix, we shall present the complete expressions 
for~$\cG_C(\vk_1,\vk_2,\vk_3)$ arrived at using these functional forms.

%%%%%%%%%%%%%%%%%%%%%%%%%%%%%%%%%%%%%%%%%%%%%%%%%%%%%%%%%%%%%%%%%%%%%%%%%%%%%%%

\subsection{Case of $p=6$}

When $p=6$, we obtain the different integrals to be
\begin{subequations}
\begin{align}
\cG_1(\vk_1,\vk_2,\vk_3)
&= -\f{i\,\HI}{32\,\Mpl\,\sqrt{k_1^3\,k_2\,k_3\, \eai}\,\ee}\,
\l(\f{\ei}{\ee}\r)^3\, (k_1^2-k_2^2-k_3^2)\nn\\
&\quad\times\biggl(\biggl\{2\, i+2\, (k_1+k_2+k3)\,\ee
+i\, \l[k_1^2-2\,k_1\,(k_2+k_3)+(k_2-k_3)^2\r]\,\ee^2\nn\\
&\quad-(k_1-k_2-k_3)\, (k_1+k_2-k_3)\, (k_1-k_2+k_3)\,\ee^3\biggr\}\,
{\mathrm{e}^{i\,(k_1+k_2+k_3)\, \ee}}\nn\\ 
&\quad+ i\,(k_1-k_2-k_3)\, (k_1+k_2-k_3)\, (k_1-k_2+k_3)\, (k_1+k_2+k_3)\,
\ee^4\; \mathrm{Ei}[i\, (k_1+k_2+k_3)\,\ee]\biggr),\\ 
\cG_2(\vk_1,\vk_2,\vk_3)
&=-\f{i\,\HI}{96\,\Mpl\,k_1\, \sqrt{k_1\,k_2\,k_3\,\eai}\,\ee}\,
\l(\f{\ei}{\ee}\r)^3\,(k_1^2-k_2^2-k_3^2)\,
\biggl(\biggl\{6\,i+6\,(k_1+k_2+k_3)\,\ee\nn\\ 
&\quad-i\, \l[k_1^2+6\, k_1\, (k_2+k_3)-3\, (k_2-k_3)^2\r]\,\ee^2
+\l[k_1^2+3\, (k_2-k_3)^2\r]\, (k_1-k_2-k_3)\,\ee^3\biggr\}\,
\mathrm{e}^{i\, (k_1+k_2+k_3)\,\ee}\nn\\ 
&\quad-i\,\l[k_1^4+2\, k_1^2\, (k_2^2+k_3^2)-3\, (k_2^2-k_3^2)^2\r]\,
\ee^4\;\mathrm{Ei}(i (k_1+k_2+k_3) \ee)\biggr),\\
\cG_3(\vk_1,\vk_2,\vk_3)
&=-\f{i\,\HI}{128\,\Mpl\, \sqrt{k_1^3\,k_2\,k_3^5\,\eai}\,\ee}\,
\l(\f{\ei}{\ee}\r)^3\,
\l[2\, k_3^2\, (k_1^2+k_2^2)+(k_1^2-k_2^2)^2-3\, k_3^4\r]\nn\\
&\quad\times\biggl(\biggl\{-6\, i-6\, (k_1+k_2+k_3)\,\ee
-i\, \l[-6\, k_3\, (k_1+k_2)+3\, (k_1-k_2)^2-k_3^2\r]\,\ee^2\nn\\
&\quad +(k_1+k_2-k_3)\,\l[3\, (k_1-k_2)^2+k_3^2\r]\,\ee^3\biggr\}\,
\mathrm{e}^{i\, (k_1+k_2+k_3)\,\ee}\nn\\ 
&\quad+i\l[2\, k_3^2\, (k_1^2+k_2^2) -3\, (k_1^2-k_2^2)^2+k_3^4\r]\,\ee^4\; 
\mathrm{Ei}[i\, (k_1+k_2+k_3)\, \ee]\biggr)+\vk_2 \leftrightarrow \vk_3,\\
\cG_4(\vk_1,\vk_2,\vk_3)
&= -\f{\HI\,\sqrt{\eai}}{16\,\Mpl\,\sqrt{k_1^7 k_2 k_3^5}\,(k_1+k_2+k_3)^5\,\ei^3}\,
\l[(k_1^2-k_2^2)^2+2\, (k_1^2+k_2^2)\, k_3^2-3\, k_3^4\r]\nn\\
&\quad\times\biggl(\biggl\{3\, \biggl[4\, k_1^5+23\, k_1^4 (k_2+k_3)
+k_1^3\, \left(46\, k_2^2+107\, k_2\, k_3+51\, k_3^2\right)\nn\\
&\quad+k_1^2\, \left(42\, k_2^3+153\, k_2^2\, k_3+178\, k_2\, k_3^2+51\, k_3^3\right)\nn\\
&\quad+k_1\, (k_2+k_3)\,
\left(18\, k_2^3+69\, k_2^2\, k_3+84\, k_2\, k_3^2+23\, k_3^3\right)
+(k_2+k_3)^2\, \left(3\, k_2^3+12\, k_2^2\, k_3+15\, k_2\, k_3^2+4\, k_3^3\right)\biggr]\nn\\
&\quad-3\, i\, (k_1+k_2+k_3)\, 
\biggl[k_1^5+8\, k_1^4\, (k_2+k_3)+k_1^3\, \left(16\, k_2^2+47\, k_2\, k_3+21\, k_3^2\right)\nn\\
&\quad+k_1^2\, (k_2+3\, k_3)\, 
\left(12\, k_2^2+27\, k_2\, k_3+7\, k_3^2\right)+k_1\, (k_2+k_3)\, 
\left(3\, k_2^3+24\, k_2^2\, k_3+39\, k_2\, k_3^2+8\, k_3^3\right)\nn\\
&\quad+k_3\, (k_2+k_3)^2\, \left(3\, k_2^2+6\, k_2\, k_3+k_3^2\right)\biggr]\,\ee
-3 \,(k_1+k_2+k_3)^2\, 
\biggl[k_1^4\, (k_2+k_3)+2\, k_1^3\, \left(k_2^2+5\, k_2\, k_3+2\, k_3^2\right)\nn\\
&\quad+k_1^2 \left(k_2^3+12\, k_2^2\, k_3+23\, k_2\, k_3^2
+4\, k_3^3\right)
+k_1\, k_3\, (k_2+k_3)\, \left(3\, k_2^2+9\, k_2\, k_3+k_3^2\right)
+k_2\, k_3^2\, (k_2+k_3)^2\biggr]\, \ee^2\nn\\
&\quad+i\, k_1\, k_3\, (k_1+k_2+k_3)^3\, 
\left[k_1^2\, (3\, k_2+k_3)+k_1\, \left(3\, k_2^2+11\, k_2\, k_3+k_3^2\right)
+3\, k_2\, k_3\, (k_2+k_3)\right]\, \ee^3\nn\\
&\quad+ k_1^2\, k_2\, k_3^2\, (k_1+k_2+k_3)^4\, \ee^4\biggr\}\,
\mathrm{e}^{i\,(k_1+ k_2+ k_3)\,\ee}
-9\,(k_1+k_2+k_3)^5\,\mathrm{Ei}[i\,(k_1+k_2+k_3)\,\ee]\biggr)\nn\\
&\quad+\vk_2 \leftrightarrow \vk_3,\\
\cG_5(\vk_1,\vk_2,\vk_3)
&=  -\f{i\HI}{256\,\Mpl\, \sqrt{k1^3\,k_2^5\,k_3^5\, \eai}\,\ee^3}\,
\l(\f{\ei}{\ee}\r)^3\,
\l[k_1^4-2\, k_1^2\, (k_2^2+ k_3^2)+ k_2^4+6\, k_2^2\, k_3^2+ k_3^4\r]\nn\\
&\quad\times\biggl(\biggl\{-24\, i-24\, (k_1+ k_2+ k_3)\,\ee
-6\, i\,\l[k_1^2-4\, k_1\, (k_2+ k_3)- k_2^2-4\, k_2\,k_3- k_3^2\r]\,\ee^2\nn\\
&\quad +2\, \l[-3\, k_3\, (k_1^2-4\, k_1\, k_2- k_2^2)
+3\, k_3^2\, (k_1+ k_2) + (k_1- k_2)^3-k_3^3\r]\,\ee^3\nn\\
&\quad+i\l[2\,k_3^3\, (k_1+ k_2)-2\, k_2\,  k_3^2\, (3\, k_1+ k_2)
-2\, k_3\, (k_1- k_2)^3+(k_1- k_2)^3\, (k_1+ k_2)- k_3^4\r]\,\ee^4\nn\\
&\quad-(k_1- k_2- k_3)\, (k_1+ k_2- k_3)\, (k_1- k_2+ k_3)\,
(k_1^2- k_2^2- k_3^2)\,\ee^5\biggr\}\, \mathrm{e}^{i\,(k_1+ k_2+ k_3)\,\ee}\nn\\
&\quad + i\,(k_1- k_2- k_3)\, (k_1+ k_2- k_3)\, (k_1- k_2+ k_3)\, 
(k_1+ k_2+ k_3)\,(k_1^2- k_2^2- k_3^2)\,\ee^6\nn\\
&\quad\times\mathrm{Ei}[i\, (k_1+ k_2+ k_3)\,\ee]\biggr),\\
\cG_6(\vk_1,\vk_2,\vk_3)
&=\f{7\,i\,\HI}{768\,\Mpl\, \sqrt{k_1^3\,k_2^5\,k_3^5\,\eai}\,\ee^3}
\l(\f{\ei}{\ee}\r)^3\,
\l[k_1^4-2\, k_1^2\, (k_2^2+k_3^2)+k_2^4+6\, k_2^2\, k_3^2+k_3^4\r]\nn\\ 
&\quad\times\biggl(3\,
\biggl\{-24\, i\, -24\, (k_1+k_2+k_3)\,\ee
+6\, i\, \l[k_1^2+4\, k_1\, (k_2+k_3)+k_2^2+4\, k_2\, k_3+k_3^2\r]\,\ee^2\nn\\ 
&\quad-2\, (k_1+k_2+k_3)\, \l[k_1^2-4\, k_1\, (k_2+k_3)
+k_2^2-4\, k_2\, k_3+k_3^2\r]\,\ee^3
-i\, \biggl[k_1^4-2\, k_1^3\, (k_2+k_3)\nn\\
&\quad+2\, k_1^2\, (k_2^2+3\, k_2\, k_3+k_3^2)
-2\, k_1\, (k_2+k_3)\, (k_2^2-4\, k_2\, k_3+k_3^2)
+(k_2-k_3)^2\,(k_2^2+k_3^2)\biggr]\,\ee^4\nn\\
&\quad+ (k_1-k_2-k_3)\, (k_1+k_2-k_3)\, (k_1-k_2+k_3)\, 
(k_1^2+k_2^2+k_3^2)\,\ee^5\biggr\}\, \mathrm{e}^{i\, (k_1+k_2+k_3)\,\ee}\nn\\ 
&\quad-i\l[-3\, (k_1^2+k_2^2)\, k_3^4\,+3\,(k_1^2-k_2^2)^2\,
(k_1^2+k_2^2) - (3\, k_1^4+2\, k_1^2\, k_2^2+3\, k_2^4)\,k_3^2
+3\, k_3^6\r]\,\ee^6\nn\\
&\quad\times\,\mathrm{Ei}[i\, (k_1+k_2+k_3)\, \ee]\biggr),
\end{align}
\end{subequations}
%%%%%%%%%%%%%%%%%%%%%%%%%%%%%%%%%%%%%%%%%%%%%%%%%%%%%%%%%%%%%%%%%%%%%%%%%%%%%%%
where $\mathrm{Ei}(x)$ denotes the exponential integral function.
We should mention that, in arriving at these results, we have regulated 
the oscillations that occur at the initial time in the sub-Hubble limit
[i.e. when $(-k\,\ei) \gg 1$] by introducing an exponential cut-off, which
is essential to single out the perturbative vacuum~\cite{Maldacena:2002vr,
Seery:2005wm,Chen:2006nt}.

%%%%%%%%%%%%%%%%%%%%%%%%%%%%%%%%%%%%%%%%%%%%%%%%%%%%%%%%%%%%%%%%%%%%%%%%%%%%%%%

\subsection{Case of $p=4$}

When $p=4$, we obtain the different integrals to be 
%%%%%%%%%%%%%%%%%%%%%%%%%%%%%%%%%%%%%%%%%%%%%%%%%%%%%%%%%%%%%%%%%%%%%%%%%%%%%%%
\begin{subequations}
\begin{align}
\cG_1(\vk_1,\vk_2,\vk_3)
&=-\f{i\,\HI}{8\,\Mpl\,\sqrt{k_1\,k_2\,k_3\,\eai}}\,
\l(\f{\ei}{\ee}\r)^2\, \l(k_1^2-k_2^2-k_3^2\r)\,
\biggl\{[1+i\,(k_1- k_2-k_3)\,\ee]\,\mathrm{e}^{i\, (k_1+k_2+k_3)\,\ee}\nn\\ 
&\quad+\,(k_1^2-k_2^2-k_3^2)\,\ee^2\, \mathrm{Ei}[i\, (k_1+k_2+k_3)\,\ee]\biggr\},\\
\cG_2(\vk_1,\vk_2,\vk_3)
&=-\f{i\,\HI}{8\,\Mpl\, \sqrt{k_1\,k_2\,k_3\, \eai}}\,
\l(\f{\ei}{\ee}\r)^2\, (k_1^2-k_2^2-k_3^2)\,
\biggl\{\biggl[1-i\,(k_1+k_2+k_3)\,\ee\nn\\
&\quad-\f{2\,k_1\, k_2\,k_3\,\ee^2}{(k_1+k_2+k_3)}\biggr]\,
\mathrm{e}^{i\, (k_1+k_2+k_3)\,\ee}
-(k_1^2+k_2^2+k_3^2)\,\ee^2\; \mathrm{Ei}[i\, (k_1+k_2+k_3)\,\ee]\biggr\},\\
\cG_3(\vk_1,\vk_2,\vk_3)
&= \f{i\,\HI}{32\,\Mpl\, \sqrt{k_1\,k_2\,k_3^5\, \eai}}\,
\l(\f{\ei}{\ee}\r)^2\,
\l[2\, k_3^2\, (k_1^2+k_2^2)+(k_1^2-k_2^2)^2-3\, k_3^4\r]\nn\\
&\quad\times\biggl\{\l[3+3\,i\, (k_1-k_2-k_3)\,\ee
-\f{2\, k_2\, k_3^2\,\ee^2}{(k_1+k_2+k_3)}\r]\,
\mathrm{e}^{i\, (k_1+k_2+k_3)\,\ee}\nn\\ 
&\quad+(3\,k_1^2-3\,k_2^2-k_3^2)\,\ee^2\; 
\mathrm{Ei}[i\, (k_1+k_2+k_3)\, \ee]\bigg\}
+\vk_2 \leftrightarrow \vk_3,\\
\cG_4(\vk_1,\vk_2,\vk_3)
&=-\frac{\HI\,\sqrt{\eai}}{16\,\Mpl\,\sqrt{k_1^5\,k_2\,k_3^5}\,
(k_1+k_2+k_3)^4\,\ei^2}\,
\l[(k_1^2-k_2^2)^2+2\, (k_1^2+k_2^2)\,k_3^2-3\, k_3^4\r]\nn\\
&\quad\times\biggl(\biggl\{-i\, \biggl[3\, k_3^2\, 
\left(9\, k_1^2+23\, k_1\, k_2+9\, k_2^2\right)
+3\, k_3\, (k_1+k_2)\, \left(5\, k_1^2+14\, k_1\, k_2+5\, k_2^2\right)\nn\\
&\quad+3\, (k_1+k_2)^2\, \left(k_1^2+3\, k_1\, k_2+k_2^2\right)
+19\, k_3^3\, (k_1+k_2)+4\, k_3^4\biggr]\nn\\
&\quad-(k_1+k_2+k_3)\,
\biggl[3\, k_3^2\, \left(3\, k_1^2+11\, k_1\, k_2+3\, k_2^2\right)
+3\, k_3\, (k_1+k_2)\, \left(k_1^2+6\, k_1\, k_2+k_2^2\right)\nn\\
&\quad+7\, k_3^3\, (k_1+k_2)+3\, k_1\, k_2\, (k_1+k_2)^2+k_3^4\biggr]\,\ee
+i\,k_3\,(k_1+k_2+k_3)^2\,
\biggl[k_3\, \left(k_1^2+8\, k_1\, k_2+k_2^2\right)\nn\\
&\quad+k_3^2\, (k_1+k_2)+3\, k_1\, k_2\, (k_1+k_2)\biggr]\,\ee^2
+k_1\, k_2\, k_3^2\, (k_1+k_2+k_3)^3\,\ee^3\biggr\}\,
\mathrm{e}^{i\,(k_1+k_2+k_3)\,\ee}\nn\\
&\quad+ 3\, i\, (k_1+k_2+k_3)^4\,\textrm{Ei}[i\,(k_1+k_2+k_3)\,\ee]\biggr)
+\vk_2\leftrightarrow\vk_3,\\
\cG_5(\vk_1,\vk_2,\vk_3)
&=-\f{i\,\HI}{128\,\Mpl\, \sqrt{k_1\,k_2^5\,k_3^5\, \eai}\,\ee^2}\,
\l(\f{\ei}{\ee}\r)^2\,
\l[k_1^4-2\, k_1^2\, (k_2^2+k_3^2)+k_2^4+6\, k_2^2\, k_3^2+k_3^4\r]\nn\\
&\quad\times\biggl(\biggl\{-18+3\,i\,\l[6\, (k_2+k_3)-2\, k_1\r]\,\ee 
+3\, \l[k_1^2-2\, k_1\, (k_2+k_3)+k_2^2+6\, k_2\, k_3+k_3^2\r]\,\ee^2\nn\\
&\quad+3\,i\,(k_1-k_2-k_3)\,(k_1+k_2-k_3)\, (k_1-k_2+k_3)\,\ee^3\biggr\}\,
\mathrm{e}^{i\, (k_1+k_2+k_3)\,\ee}\nn\\ 
&\quad+\l[2\, k_3^2\, (k_2^2-3 k_1^2) +3\, (k_1^2-k_2^2)^2+3\, k_3^4\r]\, 
\ee^4\;\mathrm{Ei}[i\, (k_1+k_2+k_3)\,\ee]\biggr),\\
\cG_6(\vk_1,\vk_2,\vk_3)
&= \f{3\,i\,\HI}{128\,\Mpl\,\sqrt{k_1\, k_2^5\,k_3^5\,\eai}\,\ee^2}\,
\l(\f{\ei}{\ee}\r)^2\,
\l[k_1^4-2\, k_1^2\, (k_2^2+k_3^2)+k_2^4+6\, k_2^2\, k_3^2+k_3^4\r]\nn\\
&\quad\times\biggl(\biggl\{-18+18\,i\, (k_1+k_2+k_3)\,\ee
-3\,\l[3\, k_1^2-6\, k_1\, (k_2+k_3)
-k_2^2-6\, k_2\, k_3-k_3^2\r]\,\ee^2\nn\\
&\quad -\f{3\,i}{(k_1+k_2+k_3)}\, 
\l[3\, k_1^4-2\, k_1^2\, (k_2^2+k_3^2)
+8\, k_1\, k_2\, k_3\, (k_2+k_3)-(k_2^2-k_3^2)^2\r]\,\ee^3\nn\\
&\quad-\f{8\, k_1\, k_2^2\, k_3^2\,\ee^4}{(k_1+k_2+k_3)}\biggr\}\,
\mathrm{e}^{i\, (k_1+k_2+k_3)\,\ee}\nn\\ 
&\quad-\, \l[9\, k_1^4-6\, k_1^2\, (k_2^2+k_3^2)-3\, k_2^4-2\, k_2^2\, k_3^2
-3\, k_3^4\r]\,\ee^4\; \mathrm{Ei}[i\, (k_1+k_2+k_3)\, \ee]\biggr).
\end{align}
\end{subequations}
%%%%%%%%%%%%%%%%%%%%%%%%%%%%%%%%%%%%%%%%%%%%%%%%%%%%%%%%%%%%%%%%%%%%%%%%%%%%%%%
We should mention that the expressions listed above have been arrived at 
without any approximations. 
In Sec.~\ref{sec:calB}, we have used these~$\cG_C(\vk_1,\vk_2,\vk_3)$ to 
evaluate the contributions~$\cB_C(\vk_1,\vk_2,\vk_3)$.
We should add that, in listing the different 
contributions~$\cB_C(\vk_1,\vk_2,\vk_3)$,
we have retained only the leading order terms in~$\ee$.
\end{widetext}

%%%%%%%%%%%%%%%%%%%%%%%%%%%%%%%%%%%%%%%%%%%%%%%%%%%%%%%%%%%%%%%%%%%%%%%%%%%%%%%

\section{Non-Gaussian contributions to~$\pb(k)$}\label{app:pc-bnl}

Since $\bnl(\vk_1,\vk_2,\vk_3)$ attains large values for $p=4$, we may 
expect a strong non-Gaussian contribution to the power spectrum $\pb(k)$
of the magnetic field.
This contribution can potentially alter the shape of the spectrum. 
In this appendix, we shall inspect such a possibility for the cases of $p=6$
and $p=4$ and values of $n$ that lead to a scale invariant power spectrum 
for the magnetic field.
Specifically, we shall arrive at an approximate estimate of the non-Gaussian
contributions.

We should mention here that similar analyses in the case of the
contributions to the scalar power spectrum~$\ps(k)$ due to the 
scalar non-Gaussianity parameter~$\fnl(\vk_1,\vk_2,\vk_3)$ has
been carried out earlier~\cite{Schmidt:2010gw,Agullo:2021oqk,
Ragavendra:2020sop,Ragavendra:2021qdu,Das:2023cum}.
We shall suitably adopt the method followed in the case of the
scalar power spectrum (in this context, see 
Refs.~\cite{Ragavendra:2021qdu,Das:2023cum}).
Using the defining relation in Eq.~\eqref{eq:bnl-def}, we can express the power 
spectrum of the magnetic field that is modified due to the non-Gaussianity 
parameter~$\bnl(\vk_1,\vk_2,\vk_3)$ to be
\begin{align}
\pbm(k) 
&= \pb(k) + \f{k^3}{4\,\pi}\,\int \d^3 {\bm q}\, 
\bnl^2(q,\vert \vk - {\bm q} \vert, k)\nn\\ 
&\times \f{\ps(q)\,\pb(\vert\vk - {\bm q}\vert)}{q^3\,\vert\vk - {\bm q}\vert^3}\nn\\
&= \pb^0\,\biggl[1 + \f{k^3}{4\,\pi}\,
\int \d^3 {\bm q}\, \bnl^2(q,\vert \vk - {\bm q} \vert, k)\nn\\ 
&\quad \times \f{\ps(q)}{q^3\,\vert\vk - {\bm q}\vert^3}\bigg].
\end{align}
In arriving at the second equation we have used the fact that $\pb(k) = \pb^0$, 
i.e. we have a scale-invariant amplitude for the power spectrum of the magnetic
field in both the cases of $p=6$ and~$4$. 
The second term in the expression within the square braces is the non-Gaussian 
contribution to the power spectrum of the magnetic field, which arises due to 
the interaction with the scalar perturbations. 
Note that the contribution is proportional to the product of the square of the 
non-Gaussianity parameter~$\bnl(\vk_1,\vk_2,\vk_3)$  and the scalar power 
spectrum~$\ps(k)$.

Recall that, in the case of $p=6$, we found that the quantity~$\bnl(\vk_1,\vk_2,\vk_3)$ 
peaks in the flattened limit [cf. Fig.~\ref{fig:bnl-2D}]. 
This behavior is similar to the orthogonal template of the scalar bispectrum 
that is often used to characterize the scalar non-Gaussianity 
parameter~$\fnl(\vk_1,\vk_2,\vk_3)$ (see, for instance, Refs.~\cite{Senatore:2009gt,
Planck:2015zfm,Planck:2019kim,Das:2023cum}).
Hence, as a reliable approximation, we can use the orthogonal template for
$\bnl(\vk_1,\vk_2,\vk_3)$ to calculate the integral that describes the
non-Gaussian contribution to $\pbm(k)$.
In other words, we shall set $\bnl(\vk_1,\vk_2,\vk_3) = \bnl^{\mathrm{ortho}}\,
\mathcal{F}^{\mathrm{ortho}}(k_1,k_2,k_3)$, where $\bnl^{\mathrm{ortho}}$ and
the function~$\mathcal{F}^{\mathrm{ortho}}(k_1,k_2,k_3)$ describes the orthogonal
shape~\cite{Senatore:2009gt,Planck:2015zfm}.
We shall set the overall amplitude to be~$\bnl^{\mathrm{ortho}} = 16$,
which is the maximum value that the non-Gaussianity parameter attains 
in the flattened limit. 
We also use the scale invariant behavior of the scalar power spectrum in this 
case, i.e. we set $\ps(k) = \ps(k_\ast)$. 
Under these conditions, the integral simplifies to 
\begin{equation}
\pbm(k) = \pb^0\,\l[1 + \ps(k_\ast)\,\f{(\bnl^{\mathrm{ortho}})^2}{4\,\pi}\, 
\mathcal{I}^{\mathrm{ortho}}\r],
\end{equation}
where~$\mathcal{I}^{\mathrm{ortho}}$ is the value of the integral performed with
the orthogonal template of the non-Gaussianity parameter and its typical value is 
$\mathcal{I}^{\mathrm{ortho}} \sim 10^{-2}$ (for the relevant calculational details 
of the integral, see Ref.~\cite{Das:2023cum}).
Using the values of $\ps(k_\ast)$, $\bnl^{\mathrm{ortho}}$ 
and~$\mathcal{I}^{\mathrm{ortho}}$, we obtain the modification to the power spectrum
of the magnetic field power due to the second term above to be
\begin{equation}
\f{\pbm(k) - \pb^0}{\pb^0} \simeq 4\times 10^{-10}.
\end{equation}
This implies that the non-Gaussian contribution to the magnetic power spectrum 
is very small in the case of $p=6$.

On the other hand, in the case of $p=4$, the quantity~$\bnl(\vk_1,\vk_2,\vk_3)$
has a similar orthogonal shape, except that there is an overall amplification 
of~$1/\ee^2$. 
This behavior of the non-Gaussianity parameter makes the $p=4$ scenario more 
interesting than the $p=6$ scenario.
While $\pb(k)=\pb^0$ in this case as well, the scalar power spectrum has a 
strong scale dependence and behaves as
\begin{equation}
\ps(k) = \ps(k_\ast)\,\l(\f{k}{k_\ast}\r)^2,
\end{equation}
where $\ps(k_\ast)$ is normalized to the CMB value at the pivot scale, as we 
discussed earlier while arriving at a suitable value of~$\HI$. 
Although $\bnl(\vk_1,\vk_2,\vk_3)$ can be approximated well by the orthogonal 
template as in the previous case, the scale dependence of~$\ps(k)$ makes the 
integral non-trivial to calculate. 
So, we shall explicitly evaluate the contribution below.

The modification to the power spectrum of the magnetic field or, in other words,
the relative difference between $\pbm(k)$ and $\pb^0$ can be expressed in terms of
$\ps(k_\ast)$ and $\bnl(\vk_1,\vk_2,\vk_3)$ as follows:
\begin{align}
\f{\pbm(k) - \pb^0}{\pb^0} 
&= \f{k^3}{4\,\pi}\,\ps(k_\ast)\,
\int \d^3 {\bm q} \l(\f{q}{k_\ast}\r)^2\nn\\ 
&\quad \times \f{\bnl^2(q,\vert \vk - {\bm q}\vert, k)}
{q^3\,\vert \vk - {\bm q}\vert^3},\nn\\
&= \f{k^3}{2\,k_\ast^2}\,\ps(k_\ast)\,\int \d \ln q\, q^2\nn\\ 
& \int \d \cos \theta\, \f{\bnl^2(q,\vert \vk - {\bm q}\vert, k)}
{\vert \vk - {\bm q}\vert^3},
\end{align}
where we have simplified the angular integral by aligning~${\bm k}$ along 
the~$\bm z$-direction in ${\bm q}$~space. 
Further, we can introduce the dimensionless variables $x = q/k$ and $y = 
\vert \vk - {\bm q} \vert/k$, so that the above expression reduces to
\begin{align}
\f{\pbm(k) - \pb^0}{\pb^0} 
&= \f{\ps(k_\ast)}{2}\,\l(\f{k}{k_\ast}\r)^2\nn\\ 
&\quad\times \int_0^\infty \d x \int_{\vert 1 - x \vert}^{1+x} \d y 
\f{\bnl^2(k\,x,k\,y,k)}{y^2}.
\end{align}

We utilize the behavior of $\bnl(\vk_1,\vk_2,\vk_3)$ peaking in the flattened 
limit to restrict the limits of the integrals in the $x$-$y$ plane. 
This is done to simplify the integration and yet obtain a reasonably good 
estimate of~$\pbm(k)$. 
Note that, while the wave number corresponding to the scalar perturbation is~$q$, 
the wave numbers of the modes of the magnetic field are~$k$ and~$\vert {\bm k} - 
{\bm q}\vert$.
Therefore, the range of~$x$ is determined by the range of~$q$ in the flattened limit.
The flattened limit corresponds to the right edge of the triangular region 
in Fig.~\ref{fig:bnl-2D}, along the line $k_2/k_3 = k_1/k_3 + 1$.
From the figure, we can see that the wave number of scalar perturbation $k_1$ (which 
is~$q$ in the integral) ranges from $k_3$ to $2\,k_3$ along this limit, where $k_3$ 
(akin to $k$ in the integral above) is the wave number corresponding to the modes 
of the magnetic field.
This sets the range of $x$ to be $[1,2]$. 
Similarly, the range of $k_2$ (equivalent to $\vert \bm k - \bm q \vert$) in this 
regime is from $0$ to $k_3$ and hence the range of $y$ corresponds to $[0, 1+x]$.
Hence, the integrals are restricted to
\begin{align}
\f{\pbm(k) - \pb^0}{\pb^0} 
&= \f{\ps(k_\ast)}{2}\,\l(\f{k}{k_\ast}\r)^2\nn\\ 
&\quad \times \int_1^2 \d x \int_0^{1+x} \d y\,
\f{\bnl^2(kx,ky,k)}{y^2}.
\end{align}
Further, we use the property that $\bnl(\vk_1,\vk_2,\vk_3) \propto 1/(k_3\,\ee)^2$ 
in this limit.
This greatly simplifies the integral, which can then be evaluated to be
\begin{align}
\f{\pbm(k) - \pb^0}{\pb^0} 
&= \f{\ps(k_\ast)}{2}\,\l(\f{k}{k_\ast}\r)^2\nn\\ 
&\quad\times \l(\f{\bnl^{\mathrm{ortho}}}{k^2\,\ee^2}\r)^2\int_1^2 
\d x \int_0^{1+x} \f{\d y}{y^2}\nn\\
& \simeq \f{\ps(k_\ast)}{2}\,\l(\f{k}{k_\ast}\r)^2\,
\l(\f{\bnl^{\mathrm{ortho}}}{k^2\,\ee^2}\r)^2 \nn \\ 
&\times \l(\f{1}{y_{\mathrm{min}}} - \ln 2 \r),
\end{align}
where we have set the lower limit of the $y$-integral to be a small finite value 
$k_{\mathrm{min}}/k = y_{\mathrm{min}} = 10^{-2}$ instead of zero, to regulate the 
divergence that arises.
Note that, in the above expression, we can set~$\bnl^{\mathrm{ortho}} \simeq 10$, 
which is roughly the amplitude of the parameter in the flattened limit apart from 
the $1/(k^2\,\ee^2)$ factor, as can be seen from Fig.~\ref{fig:bnl-2D}.

It should be clear from the above expression that the relative difference is 
much larger than unity, mainly because of the presence of the factor~$1/(k\,\ee)^4$. 
Therefore, the non-Gaussian contribution to $\pb(k)$ dominates over $\pb^0$ 
in the case of $p=4$. 
Such a modified $\pbm(k)$ can be expressed as
\begin{align}
\pbm(k)  &\simeq  \f{\pb^0\,\ps(k_\ast)}{2\,y_{\mathrm{min}}}\,
\f{(\bnl^{\mathrm{ortho}})^2}{k^2\,k_\ast^2\,\ee^4}\nn\\
&\simeq 1.5 \times 10^{-63}\,\l(\f{k_\ast}{k}\r)^2\Mpl^4.
\end{align}
Note that $\pb^0 \simeq 9\,\HI^4/(4\,\pi^2) \simeq 8.9 \times 10^{-140}\,\Mpl^4$.
So, it is clear that $\pbm(k)$ is dominated by the non-Gaussian contribution.
Moreover, $\bnl(\vk_1,\vk_2,\vk_3)$ introduces a strong scale dependence as 
$\pbm(k) \propto 1/k^2$, unlike the strictly scale invariant $\pb^0$.
While, the overall amplitude is still much smaller than required around $k_\ast$, 
it can lead to large contributions over larger scales such that $k \ll k_\ast$.

%%%%%%%%%%%%%%%%%%%%%%%%%%%%%%%%%%%%%%%%%%%%%%%%%%%%%%%%%%%%%%%%%%%%%%%%%%%%%%%

\section{Non-Gaussianity parameter in other cases}\label{app:DC}

\begin{figure*}
\centering
\includegraphics[width=0.475\textwidth]{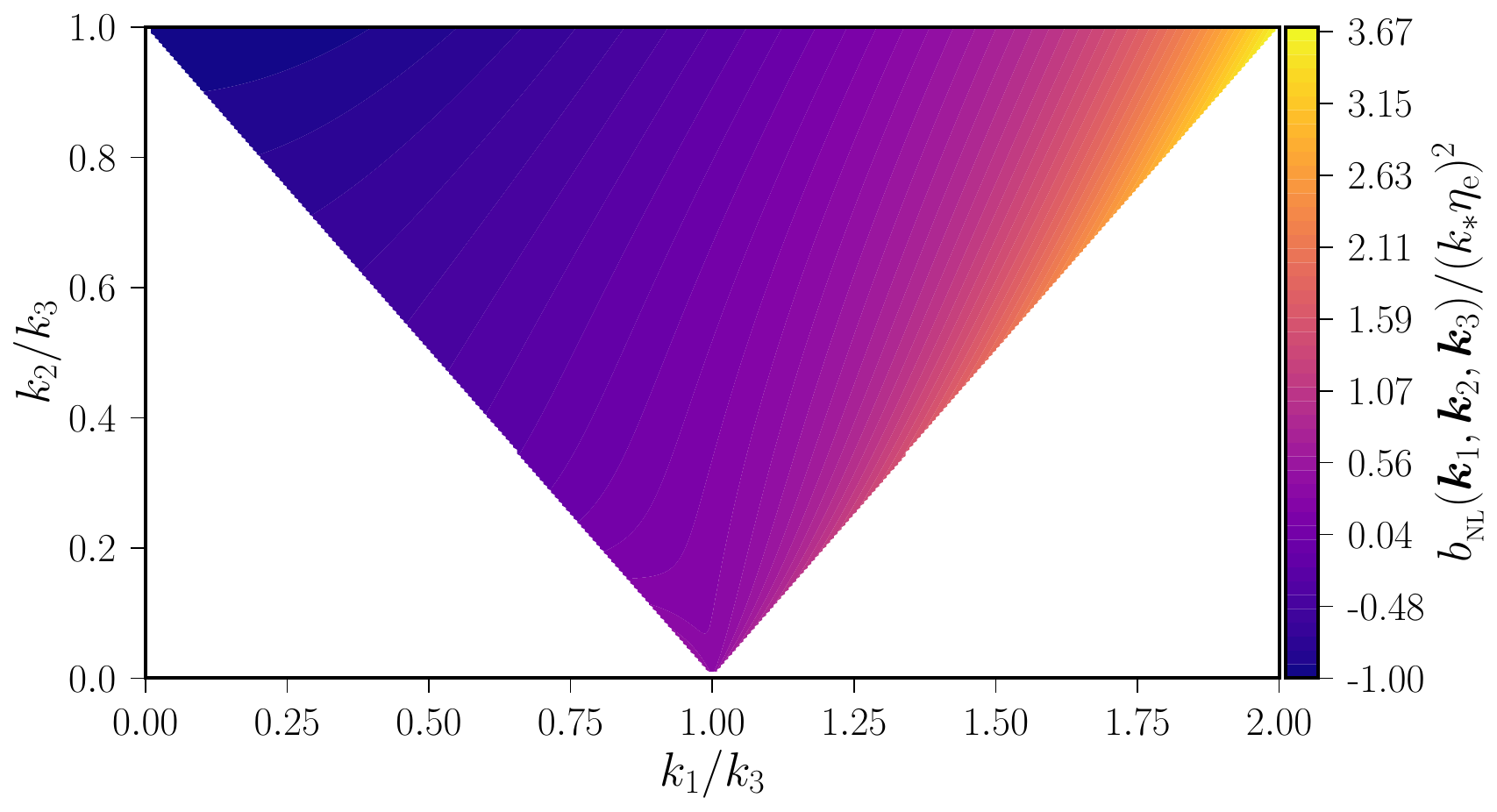}
\includegraphics[width=0.475\textwidth]{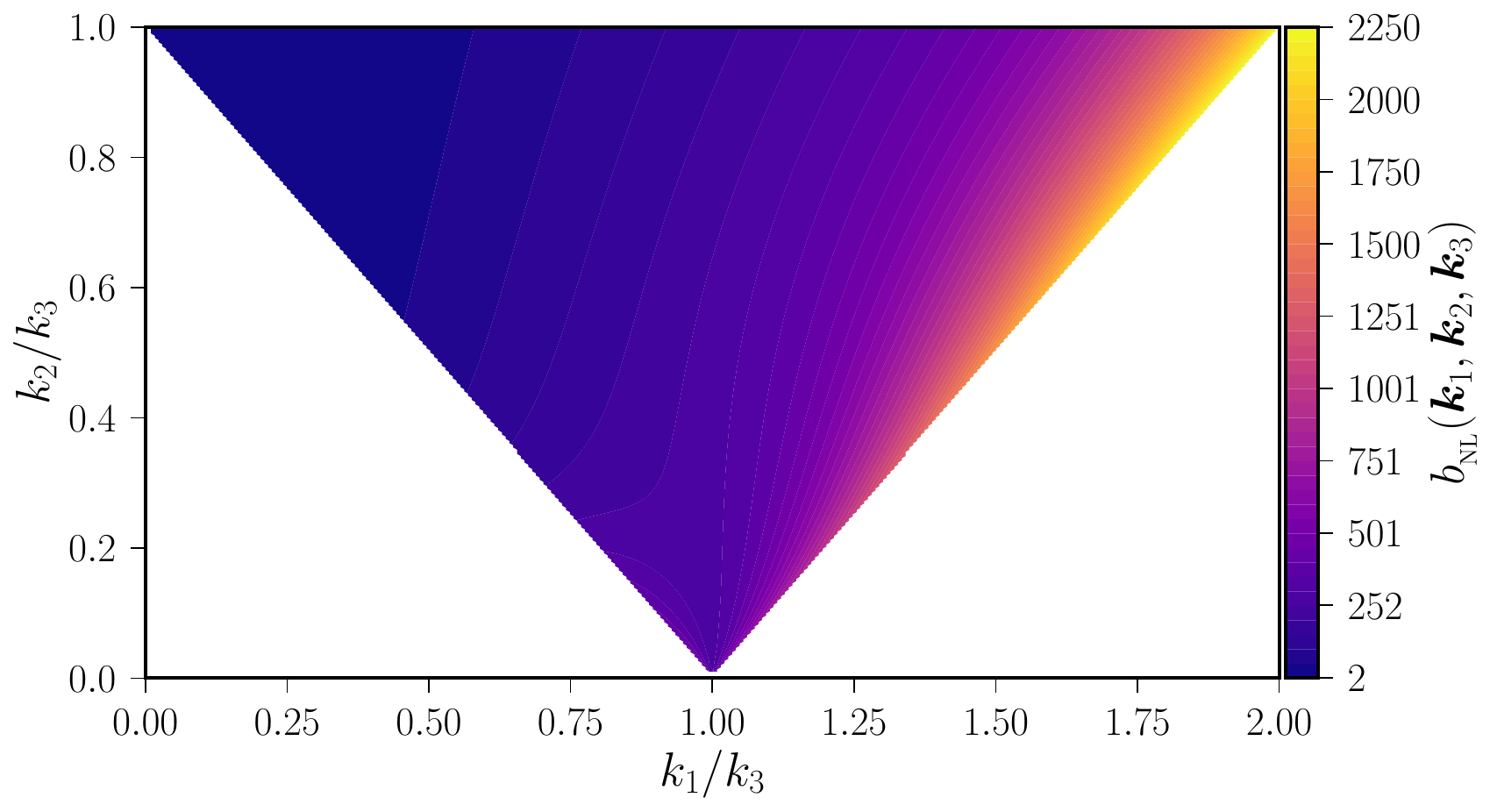}
\caption{Density plots of the non-Gaussianity parameter~$\bnl(\vk_1,\vk_2,\vk_3)$
in the ultra slow roll scenario with $p=6$ and $n=-1/3$ (on the left) and slow 
roll inflation (on the right).
In the ultra slow roll case, we have chosen the parameter values to be $\eai 
= 10^{-3}$, $\ei = -10^6\,\mathrm{Mpc}$ and $\ee = -10^{-19}\,\mathrm{Mpc}$,
as in the previous figure.
To arrive at the results in the slow roll case, we have assumed the non-conformal
coupling function to be of the form $J(\phi)$, as in the action in Eq.~\eqref{eq:ema-1}.
Moreover, we have assumed that $J\propto a^2$, corresponding to $\bar{n}=2$,
which leads to a scale invariant power spectrum for the magnetic field.
Also, we have chosen the values of the parameters to be $\eai = 10^{-3}$, 
$\HI=1.25 \times 10^{-5}\,\Mpl$ and $\ee = -10^{-19}\,\mathrm{Mpc}$. 
Lastly, we have set $k_3$ to be the pivot scale $k_\ast$ in both the 
cases.}\label{fig:bnl-ac}
\end{figure*}

To further investigate the validity of the consistency relation given 
in Eq.~\eqref{eq:bnl-cr}, let us consider the ultra slow roll scenario 
wherein $p=6$ with $n=-1/3$.
In this case, since $n\, p=-2$, the power spectrum of the magnetic field
is not scale invariant but instead has a strong blue tilt and behaves 
as~$k^2$ (i.e. $\nb=2$). 
Moreover, in this case, we find that the contributions to the cross-correlation
from the terms~$\cB_1(\vk_1,\vk_2,\vk_3)$, $\cB_2(\vk_1,\vk_2,\vk_3)$,  
$\cB_3(\vk_1,\vk_2,\vk_3)$, $\cB_5(\vk_1,\vk_2,\vk_3)$, and $\cB_6(\vk_1,\vk_2,\vk_3)$ 
are of the same order.
At the leading order, the resulting non-Gaussianity parameter can be
obtained to be
\begin{align}
\bnl(\vk_1,\vk_2,\vk_3) 
&= \f{\ee^2}{12\, k_2^2\, k_3^2\, (k_2 + k_3)}\,
\biggl[k_1^7\nn\\ 
&\quad+6\, k_1^2\, k_2^2\, k_3^2\, (k_2+k_3)- 2\, k_1^5\, (k_2^2+k_3^2)\nn\\ 
&\quad - 6\, k_2^2\, k_3^2\, (k_2+k_3)\,
(k_2^2 + k_3^2) \nn\\ 
&\quad+ k_1^3\, (k_2^4 + 6\, k_2^2\, k_3^2 
+ k_3^4)\biggr].\label{eq:bnl_p6_nm1by3}
\end{align}
In Fig.~\ref{fig:bnl-ac}, we have plotted the above non-Gaussianity parameter 
as a density plot for an arbitrary configuration of wave vectors, in the same 
manner as we have done earlier.
%%%%%%%%%%%%%%%%%%%%%%%%%%%%%%%%%%%%%%%%%%%%%%%%%%%%%%%%%%%%%%%%%%%%%%%%%
%%%%%%%%%%%%%%%%%%%%%%%%%%%%%%%%%%%%%%%%%%%%%%%%%%%%%%%%%%%%%%%%%%%%%%%%%
Since the absolute values of the numbers involved are very small, we have
divided the quantity $\bnl(\vk_1,\vk_2,\vk_3)$ by $(k_\ast^2\, \ee^2)$, where
$k_\ast=0.05\,\mpcinv$, i.e. the pivot scale. 
When $(-k_\ast\,\ee) \ll 1$, in the equilateral, flattened and squeezed 
limits, we find that the non-Gaussianity parameter reduces to
%%%%%%%%%%%%%%%%%%%%%%%%%%%%%%%%%%%%%%%%%%%%%%%%%%%%%%%%%%%%%%%%%%%%%%%%%
\begin{subequations}
\begin{align}
\f{\bnl^{\mathrm{eq}}(k_\ast)}{k_\ast^2\, \ee^2} 
&= -\f{7}{24} = -0.29,\\
\f{\bnl^{\mathrm{fl}}(k_\ast)}{k_\ast^2\, \ee^2} 
&= \f{11}{3} = 3.67,\\
\f{\bnl^{\mathrm{sq}}(k_\ast)}{k_\ast^2\, \ee^2} 
&\simeq -1.\label{eq:bnl-app-sq}
\end{align}
\end{subequations}
%%%%%%%%%%%%%%%%%%%%%%%%%%%%%%%%%%%%%%%%%%%%%%%%%%%%%%%%%%%%%%%%%%%%%%%%%
These numbers can be matched with the plot in Fig.~\ref{fig:bnl-ac}.
In this case, it is important to note that, contrary to the previous two 
cases, the values of $\bnl(\vk_1,\vk_2,\vk_3)$ are very small.
Interestingly, the value is smaller in the equilateral limit than in the 
squeezed limit. 
Moreover, the non-Gaussianity parameter does not vanish in the squeezed
limit (though it proves to be rather small). 

Let us now compare the $\bnl(\vk_1,\vk_2,\vk_3)$ we have obtained in the 
ultra slow roll scenarios of interest with the results arrived at in the
standard slow roll inflationary scenario.
In the latter case, as is often done, we shall assume that the
electromagnetic field is described by the action in Eq.~\eqref{eq:ema-1} 
with a non-conformal coupling function of the form~$J(\phi)$.
For the slow roll case, in a background that is approximated to be that
of de Sitter, and with a non-conformal coupling function that behaves 
as $J \propto a^2$ [i.e. when $\bar{n}=2$, cf. Eqs.~\eqref{eq:Ja} 
and~\eqref{eq:Je}], we obtain a scale invariant power spectrum for the 
magnetic field (i.e. $\nb=0$). 
Also, in this case, the three-point cross-correlation of interest and the
corresponding non-Gaussianity parameter can be easily calculated (for
details, see Refs.~\cite{Jain:2012ga,Nandi:2021lpf}).
For suitable choices of parameters such that the inflationary scalar 
power spectrum is normalized to the value suggested by the CMB (say, 
when $\epsilon_{1}\simeq 10^{-3}$ and $\ee=-10^{-19}\,\mathrm{Mpc}$), we 
obtain the following estimates for $\bnl(\vk_1,\vk_2,\vk_3)$ in the 
various limits of wave numbers (when evaluated at the pivot scale):
\begin{subequations}\label{eq:bnl-dS-n2}
\begin{align}
\bnl^{\mathrm{eq}}(k_\ast) &\simeq 180.18,\\
\bnl^{\mathrm{fl}}(k_\ast) &\simeq 2249.50,\\
\bnl^{\mathrm{sq}}(k_\ast) &= 2.
\end{align}
\end{subequations}
Note that in order to arrive at these numbers, we have neglected the 
contribution due to the term $\cB_4(\vk_1,\vk_2,\vk_3)$, since this 
term, being directly proportional to the first slow roll 
parameter~$\epsilon_1$, would be subdominant compared to the other terms.
These numbers can be matched with the density plot in Fig.~\ref{fig:bnl-ac}, 
where we have plotted $\bnl(\vk_1,\vk_2,\vk_3)$ for an arbitrary triangular
configuration of wave vectors. 
It is also clear from the above-mentioned value of $\bnl^{\mathrm{sq}}(k)$ 
that the consistency relation in Eq.~\eqref{eq:bnl-cr} is satisfied in this 
case (since $\nb=0$).

Finally, if we consider another slow roll case in an approximately de Sitter 
background with $J(\phi) \propto a$ (corresponding to $\bar{n}=1$), we obtain 
that $\pb(k) \propto k^2$ [i.e. $\nb=2$, cf. Eq.~\eqref{eq:nb}]. 
Evaluating the three-point function in this scenario leads to the following 
values of the non-Gaussianity parameter $\bnl(\vk_1,\vk_2,\vk_3)$ in the 
different limits of wave numbers (when $\epsilon_{1}\simeq 10^{-3}$ and 
$\ee=-10^{-19}\,\mathrm{Mpc}$):
\begin{subequations}\label{eq:bnl-dS-n1}
\begin{align}
\bnl^{\mathrm{eq}}(k) &\simeq 0.96,\\
\bnl^{\mathrm{fl}}(k) &\simeq 3,\\
\bnl^{\mathrm{sq}}(k) &= 1.
\end{align}
\end{subequations}
As in the preceding case, it is evident from the above value 
of $\bnl^{\mathrm{sq}}(k)$ that the consistency relation in Eq.~\eqref{eq:bnl-cr} 
is satisfied in this case as well.
%%%%%%%%%%%%%%%%%%%%%%%%%%%%%%%%%%%%%%%%%%%%%%%%%%%%%%%%%%%%%%%%%%%%%%%%%

%%%%%%%%%%%%%%%%%%%%%%%%%%%%%%%%%%%%%%%%%%%%%%%%%%%%%%%%%%%%%%%%%%%%%%%%%
\bibliographystyle{apsrev4-2}
\bibliography{prd_v2}
%%%%%%%%%%%%%%%%%%%%%%%%%%%%%%%%%%%%%%%%%%%%%%%%%%%%%%%%%%%%%%%%%%%%%%%%%
\end{document}